\input epsf
\documentstyle{amsppt}
\overfullrule=0pt
\newcount\mgnf\newcount\tipi\newcount\tipoformule\newcount\greco
\tipi=2          %uso caratteri: 2=cmcompleti, 1=cmparziali, 0=amparziali
\tipoformule=0   %=0 da numeroparagrafo.numeroformula; se no numero
                 %assoluto

\global\newcount\numsec\global\newcount\numfor
\global\newcount\numapp\global\newcount\numcap
\global\newcount\numfig\global\newcount\numpag
\global\newcount\numnf
\global\newcount\numtheo

\def\SIA #1,#2,#3 {\senondefinito{#1#2}%
\expandafter\xdef\csname #1#2\endcsname{#3}\else
\write16{???? ma #1,#2 e' gia' stato definito !!!!} \fi}

\def \FU(#1)#2{\SIA fu,#1,#2 }

\def\etichetta(#1){(\veroparagrafo.\veraformula)%
\SIA e,#1,(\veroparagrafo.\veraformula) %
\global\advance\numfor by 1%
\write15{\string\FU (#1){\equ(#1)}}%
\write16{ EA #1 ==> \equ(#1)  }}

\def\etichettat(#1){\veroparagrafo.\veratheorema:%
\SIA e,#1,{\veroparagrafo.\veratheorema} %
\global\advance\numtheo by 1%
\write15{\string\FU (#1){\thu(#1)}}%
\write16{ TtH #1 ==> \thu(#1)  }}

\def\etichettaa(#1){(A\veraappendice.\veraformula)
 \SIA e,#1,(A\veraappendice.\veraformula)
 \global\advance\numfor by 1
 \write15{\string\FU (#1){\equ(#1)}}
 \write16{ EA #1 ==> \equ(#1) }}
\def\getichetta(#1){Fig. \verafigura
 \SIA g,#1,{\verafigura}
 \global\advance\numfig by 1
 \write15{\string\FU (#1){\graf(#1)}}
 \write16{ Fig. #1 ==> \graf(#1) }}
\def\retichetta(#1){\numpag=\pgn\SIA r,#1,{\verapagina}
 \write15{\string\FU (#1){\rif(#1)}}
 \write16{\rif(#1) ha symbol  #1  }}
\def\etichettan(#1){(n\verocapitolo.\veranformula)
 \SIA e,#1,(n\verocapitolo.\veranformula)
 \global\advance\numnf by 1
\write16{\equ(#1) <= #1  }}

\newdimen\gwidth
\gdef\profonditastruttura{\dp\strutbox}
\def\senondefinito#1{\expandafter\ifx\csname#1\endcsname\relax}
\def\BOZZA{
\def\alato(##1){
 {\vtop to \profonditastruttura{\baselineskip
 \profonditastruttura\vss
 \rlap{\kern-\hsize\kern-1.2truecm{$\scriptstyle##1$}}}}}
\def\galato(##1){ \gwidth=\hsize \divide\gwidth by 2
 {\vtop to \profonditastruttura{\baselineskip
 \profonditastruttura\vss
 \rlap{\kern-\gwidth\kern-1.2truecm{$\scriptstyle##1$}}}}}
\def\verapagina{
{\romannumeral\number\numcap}.\number\numsec.\number\numpag}}

\def\alato(#1){}
\def\galato(#1){}
\def\veroparagrafo{\number\numsec}\def\veraformula{\number\numfor}
\def\veraappendice{\number\numapp}
\def\verapagina{\number\pageno}\def\veranformula{\number\numnf}
\def\verafigura{{\romannumeral\number\numcap}.\number\numfig}
\def\verocapitolo{\number\numcap}\def\veranformula{\number\numnf}
\def\veratheorema{\number\numtheo}
\def\Eqn(#1){\eqno{\etichettan(#1)\alato(#1)}}
\def\eqn(#1){\etichettan(#1)\alato(#1)}
\def\TH(#1){{\etichettat(#1)\alato(#1)}}%\csname fu#1\endcsname\fi}
\def\thv(#1){\senondefinito{fu#1}$\clubsuit$#1\else\csname fu#1\endcsname\fi}
\def\thu(#1){\senondefinito{e#1}\thv(#1)\else\csname e#1\endcsname\fi}

\def\Eq(#1){\eqno{\etichetta(#1)\alato(#1)}}
\def\eq(#1){\etichetta(#1)\alato(#1)}
\def\Eqa(#1){\eqno{\etichettaa(#1)\alato(#1)}}
\def\eqa(#1){\etichettaa(#1)\alato(#1)}
\def\dgraf(#1){\getichetta(#1)\galato(#1)}
\def\drif(#1){\retichetta(#1)}

\def\eqv(#1){\senondefinito{fu#1}$\clubsuit$#1\else\csname fu#1\endcsname\fi}
\def\equ(#1){\senondefinito{e#1}\eqv(#1)\else\csname e#1\endcsname\fi}
\def\graf(#1){\senondefinito{g#1}\eqv(#1)\else\csname g#1\endcsname\fi}
\def\rif(#1){\senondefinito{r#1}\eqv(#1)\else\csname r#1\endcsname\fi}
%%%%%%%%%%%%%%%%%%%%%%%%%%%%%%%%%%%%%%%%%%%%%%%%%%%%%%%%%%%%%%
%%%%%%%%%%%%%%%%%% Numerazione verso il futuro ed eventuali paragrafi
%%%%%%%%%%%%%%%%%% precedenti non inseriti nella scheda da compilare
%%%%%%%%%%%%%%%%%% e elenco referenze bibliografiche creato in
%%%%%%%%%%%%%%%%%% \jobname.bib
\def\bib[#1]{[#1]\numpag=\pgn
\write13{\string[#1],\verapagina}}

\def\include#1{
\openin13=#1.aux \ifeof13 \relax \else
\input #1.aux \closein13 \fi}

\openin14=\jobname.aux \ifeof14 \relax \else
\input \jobname.aux \closein14 \fi
\openout15=\jobname.aux%\write15
\openout13=\jobname.bib
%%%%%%%%%%%%%%%%%%%%%%%%%%%%

%%%%%%%%%%%%%%%%%%%%%%%%%%%%%%%%%%%%%%%%%%%%%%%%%%%%%%%%%%%%%%

\ifnum\tipoformule=1\let\Eq=\eqno\def\eq{}\let\Eqa=\eqno\def\eqa{}
\def\equ{}\fi

%%%%%%%%%%%%%%%%%%%%%%%%%%%%%%%%%%%%%%%%%%%%%%
%%%%%%%%%%%%%%%%%%%%%  Numerazione pagine

{\count255=\time\divide\count255 by 60 \xdef\hourmin{\number\count255}
        \multiply\count255 by-60\advance\count255 by\time
   \xdef\hourmin{\hourmin:\ifnum\count255<10 0\fi\the\count255}}

\def\oramin{\hourmin }

\def\data{\number\day/\ifcase\month\or january \or february \or march \or
april \or may \or june \or july \or august \or september
\or october \or november \or december \fi/\number\year;\ \oramin}

\def\titdate{ \ifcase\month\or January \or February \or March \or
April \or May \or June \or July \or August \or September
\or October \or November \or December \fi \number\day, \number\year;\ \oramin}
% Date to be used for title page

\setbox200\hbox{$\scriptscriptstyle \data $}

\newcount\pgn \pgn=1
\def\foglio{\number\numsec:\number\pgn
\global\advance\pgn by 1}
\def\foglioa{A\number\numsec:\number\pgn
\global\advance\pgn by 1}

\footline={\rlap{\hbox{\copy200}}\hss\tenrm\folio\hss}
%\footline={\hss\tenrm\folio\hss}

%%%%%%%%

\global\newcount\numpunt

\magnification=\magstephalf
%\magnification=1200
\baselineskip=16pt
\parskip=8pt

\voffset=2.5truepc
\hoffset=0.5truepc
\hsize=6.1truein
\vsize=8.4truein %(including running head)
%\vsize=17truecm
%\hsize=11.5truecm
{\headline={\ifodd\pageno\rightheadline \else \leftheadline \fi}}
\def\rightheadline{\it  {tralala}\hfil\tenrm\folio}
\def\leftheadline{\tenrm \folio \hfil\it  {Section $\ver$}}

\def\a{\alpha}

\def\d{\delta}
\def\e{\epsilon}

\def\f{\phi}

\def\s{\sigma}

\def\L{\Lambda}
\def\G{\Gamma}
\def\O{\Omega}
\def\S{\Sigma}

\def\1{{1\kern-.25em\roman{I}}}
\def\eu{{1\kern-.25em\roman{I}}}
\def\f1{{1\kern-.25em\roman{I}}}

\def\R{{\Bbb R}}  %%
\def\N{{\Bbb N}}  %%
\def\P{{\Bbb P}}  %% carateri piu belle per campi di nombre
\def\Z{{\Bbb Z}}  %%
\def\E{{\Bbb E}}  %%
  %%

  %%

%\def\P{\hskip.2em\hbox{\rm P\kern-0.8em{I}\hskip.7em}}

% Spezielle Definitionen

\let\cal=\Cal
\def\AA{{\cal A}}
\def\BB{{\cal B}}
\def\CC{{\cal C}}
\def\DD{{\cal D}}

\def\FF{{\cal F}}

\def\LL{{\cal L}}

\def\PP{{\cal P}}
\def\QQ{{\cal Q}}
\def\RR{{\cal R}}
\def\SS{{\cal S}}

\def\A{{\cal A}}

\def\chap #1#2{\line{\ch #1\hfill}\numsec=#2\numfor=1\numtheo=1}

\def\limlaw{\buildrel \DD\over\rightarrow}

%   Non-character macros

%\newcount\foot
%\foot=1
%\def\note#1{\footnote{${}^{\number\foot}$}{\ftn #1}\advance\foot by 1}
%\def\note#1{\plainfootnote{${}^{\number\foot}$}{\ftn #1}\advance\foot\by 1}
\def\note#1{\footnote{#1}}

\def\frac#1#2{{#1\over #2}}
\def\sfrac#1#2{{\textstyle{#1\over #2}}}

\def\text#1{\quad{\hbox{#1}}\quad}
\def\newpage{\vfill\eject}
\def\proposition #1{\noindent{\thbf Proposition #1}}

\def\theo #1{\noindent{\thbf Theorem {#1} }}

\def\proof{{\noindent\pr Proof: }}

\def\endproof{$\diamondsuit$}
\def\remark{\noindent{\bf Remark: }}
\def\thanks{\noindent{\bf Acknowledgements: }}

\font\pr=cmbxsl10

%\font\thbf=cmcsc10 scaled\magstep1
\font\thbf=cmbxsl10 scaled\magstephalf
% Font-Definitions

\font\ch=cmbx12
\font\ftn=cmr8

\font\it=cmti10
\font\bf=cmbx10
\font\sm=cmr7

\newfam\msafam
\newfam\msbfam
\newfam\eufmfam
%
% -------------------------------------------------- math macros --------
%
%
\def\hexnumber#1{%
  \ifcase#1 0\or 1\or 2\or 3\or 4\or 5\or 6\or 7\or 8\or
  9\or A\or B\or C\or D\or E\or F\fi}
\font\tenmsa=msam10
\font\sevenmsa=msam7
\font\fivemsa=msam5
\textfont\msafam=\tenmsa
\scriptfont\msafam=\sevenmsa
\scriptscriptfont\msafam=\fivemsa
\edef\msafamhexnumber{\hexnumber\msafam}%
%
%   \mathchardef\restriction"1\msafamhexnumber16
%   "class, family, position (found on amstex guide)
%
\mathchardef\restriction"1\msafamhexnumber16
\mathchardef\ssim"0218
\mathchardef\square"0\msafamhexnumber03
\mathchardef\eqd"3\msafamhexnumber2C
\def\QED{\ifhmode\unskip\nobreak\fi\quad
  \ifmmode\square\else$\square$\fi}
\font\tenmsb=msbm10
\font\sevenmsb=msbm7
\font\fivemsb=msbm5
\textfont\msbfam=\tenmsb
\scriptfont\msbfam=\sevenmsb
\scriptscriptfont\msbfam=\fivemsb
\def\Bbb#1{\fam\msbfam\relax#1}
\font\teneufm=eufm10
\font\seveneufm=eufm7
\font\fiveeufm=eufm5
\textfont\eufmfam=\teneufm
\scriptfont\eufmfam=\seveneufm
\scriptscriptfont\eufmfam=\fiveeufm

\def\({\left(}
\def\){\right)}
%
%-------------------------------------------------------------------
%
% ------- Per compatibilita'
%

\let\Z=\integer

%
%\numsec=1\numfor=1

\def\rank{\hbox{\rm rank}}
\def\cov{\hbox{\rm cov}}

{\headline={\ifodd\pageno\rightheadline \else \leftheadline \fi}}
\def\rightheadline{\it  {Mertens universality}\hfil\tenrm\folio}
\def\leftheadline{\tenrm \folio \hfil\it  {Section $\ver$}}

\font\tit=cmbx12
\font\aut=cmbx12
\font\aff=cmsl12
\overfullrule=0pt
\def\s{\char'31}
%\nopagenumbers
%{$  $}
%\vskip1.5truecm
\centerline{\tit Local energy statistics in disordered systems:}
\vskip.1truecm
\centerline{\tit  a
proof of the local REM conjecture
 \note{Research supported in part by the
DFG in the Dutch-German Bilateral Research Group ``Mathematics of
Random Spatial Models from Physics and Biology'' and by the
European Science Foundation in the Programme RDSES.} }
\vskip.2truecm
%\vskip.2truecm
%\centerline{\tit }
\vskip0.5truecm
%\centerline{\tit }
%\vskip2.5cm
%\input epsf
\centerline{\aut Anton Bovier \note{ e-mail:
bovier\@wias-berlin.de}  } \vskip.1truecm \centerline{\aff
Weierstra\s {}--Institut} \centerline{\aff f\"ur Angewandte
Analysis und Stochastik} \centerline{\aff Mohrenstrasse 39, 10117
Berlin, Germany} \centerline{and} \centerline{\aff Institut f\"ur
Mathematik} \centerline{\aff Technische Universit\"at Berlin}
\centerline{\aff Strasse des 17. Juni 136,
%\centerline{\aff
12623 Berlin, Germany}
\vskip.4truecm
\centerline{\aut  Irina Kurkova\note{\ftn
e-mail: kourkova\@ccr.jussieu.fr}}
\vskip.1truecm
\centerline{\aff Laboratoire de Probabilit\'es et Mod\`eles
Al\'eatoires}
\centerline{\aff Universit\'e Paris 6}
\centerline{\aff 4, place Jussieu, B.C. 188}
\centerline{\aff 75252 Paris, Cedex 5, France}

\vskip0.2truecm\rm
\def\s{\sigma}
\noindent {\bf Abstract:} Recently, Bauke and Mertens conjectured that
the local statistics of energies in random spin systems with discrete
spin space should in most circumstances be the same as in the random
energy model. Here we give necessary conditions for this hypothesis
to be true, which we show to hold in wide classes of examples: short
range spin glasses and mean field spin glasses of the SK type. We also
show that, under certain conditions, the conjecture holds even if
energy levels that grow moderately
 with the volume of the system are considered.
%Finally, we show that the conjecture fails if levels near the extremes
%are considered, and show that a complex and rich structure exmerges in
%this
%case in the example of Derrida's generalized random energy models.



{\headline={\ifodd\pageno\rightheadline \else \leftheadline \fi}}
\def\rightheadline{\it  {Local REM  conjecture}\hfil\tenrm\folio}
\def\leftheadline{\tenrm \folio \hfil\it  {Section $\ver$}}

\newpage

\chap{1. Introduction.}1
In a recent paper \cite{BaMe}, Bauke and Mertens have formulated an
interesting conjecture regarding the behaviour of local energy level
statistics in disordered systems. Roughly speaking, their conjecture
can be formulated as follows. Consider a random Hamiltonian,
$H_N(\s)$, i.e.,  a real-valued random function on some  product space,
$\SS^{\L_N}$,
where $\SS$ is a finite space, typically $\{-1,1\}$, of the form
$$
H_N(\s) =\sum_{A\subset \L_N} \Phi_A(\s),
\Eq(Mer.1)
$$
  where $\L_N$ are finite subsets of $Z^{d}$ of cardinality, say,
$N$.
 The sum runs over
  subsets, $A$, of $\L_N$
 and $\Phi_A$ are
random local functions, typically of the form
$$
\Phi_A(\s) =J_A\prod_{x\in A}\s_x
\Eq(Mer.2)
$$
where $J_A$, $A\subset \Z^d$, is a family of (typically independent)
random variables, defined on some probability space,
$(\O,\FF,\P)$, whose distribution is not too singular.
%We may assume for simplicity that $\E H_N(\s)=0$.
In such a
situation, for typical $\s$, $H_N(\s) \sim \sqrt N$, while
$\sup_{\s}H_N(\s)\sim N$. Bauke and Mertens then ask the following
question:
Given a fixed number, $E$, what is the statistics of the values
$N^{-1/2}H_N(\s)$ that are closest to this number, and how are
configurations, $\s$, for which these good approximants of $E$ are
realised,
distributed on $\SS^{\L_N}$? Their conjectured answer, which at first
glance seems rather surprising, is quite  simple:
find $\delta_{N, E}$ such
that $\P( |N^{-1/2}H_N(\s)-E|\leq  b \delta_{N, E}) \sim |\SS|^{-N} b$; then,
 the
collection of points, $ \delta_{N, E}^{-1} |N^{-1/2}H_N(\s)-E|$, over
     all  $\s\in \SS^{\L_N}$,
converges to a Poisson point process on $\R_+$.
   Furthermore,  for any finite
$k$, the $k$-tuple of configurations,
   $\s^1,\s^2,\dots, \s^k$, where the $k$-best approximations
are realised, is such that all of its elements have maximal Hamming
distance between each other. In other words, the asymptotic behavior of these
best approximants of $E$ is the same, as if the random variables
$H_N(\s)$ were all independent Gaussian random
  variables with variance $N$, i.e., as if we were dealing with the
random energy model (REM) \cite{Der1}. Bauke and Mertens call
this ``universal REM like behaviour''.

A comparable result had previously been conjectured by Mertens
\cite{Mer1} in the particular case of the {\it number partitioning
  problem}. In that case, the function $H_N$ is simply given by
$$
H_N(\s)= \sum_{i=1}^N X_i\s_i,
\Eq(Mer.3)
$$
with $X_i$ i.i.d.\ random variables uniformly distributed on $[0,1]$,
 $\s_i\in \{-1,1\}$, and one is interested in the distribution of
energies near the value zero (which corresponds to an optimal
partitioning of the $N$ random variables, $X_i$, into two groups such
that their sum in each group is as similar as possible).
This conjecture was later proven by Borgs, Chayes, and Pittel
\cite{BCP, BCMP}. It should be noted that in this problem, one needs, of
course,   take
care of the obvious symmetry of the Hamiltonian under the
transformation $\s\rightarrow -\s$.  An extension of these results in
the spirit of the REM conjecture was proven recently in \cite{BCMN},
i.e., when the value zero is replaced by an arbitrary value, $E$.

In \cite{BK2} we generalised this result to the case of the
$k$-partitioning problem, where the random function to be
considered is actually vector-valued (consisting of the vector of
differences between the sums of the random variables in each of
the $k$ subsets of the partition). To be precise, we considered
the special case where the subsets of the partition are required
to have the same cardinality, $N/k$ (restricted $k$-partitioning
problem). The general approach to the proof we developed in that
paper sets the path towards the proof of the conjecture by Bauke
and Mertens that we will present here.

The universality conjecture suggests that correlations are
irrelevant for the properties of the local energy statistics of
disordered systems for energies near ``typical energies''. On the
other hand, we know that correlations must play a r\^ole for the
extremal energies near the maximum of $H_N(\s)$. Thus, there are
two questions beyond the original conjecture that naturally pose
themselves: (i) assume we consider instead of fixed $E$,
$N$-dependent energy levels, say, $E_N=N^{\a} C$. How fast can we
allow $E_N$ to grow for the REM-like behaviour to hold? and (ii)
what type of behaviour can we expect once $E_N$ grows faster than
this value?  We will see that the answer to the first question
depends on the properties of $H_N$, and we will give an answer in
models with Gaussian couplings. The answer to question (ii)
requires a detailed understanding of $H_N(\s)$ as a random
process, and we will be able to give a complete answer on only in
the case of the GREM, when $H_N$ is a hierarchically correlated
Gaussian process. This will be discussed in a separate paper
\cite{BK05}.

Our paper will be organized as follows.
 In Chapter 2, we prove an abstract
theorem, that implies the REM-like-conjecture under three hypothesis.
This will give us some heuristic understanding why and when such a
conjecture should be true. In Chapter 3 we then show that the hypothesis
of the theorem are fulfilled in two classes of examples:
  $p$-spin Sherrington-Kirkpatrick like models and
short range Ising
models on the lattice.
In both cases we establish conditions on how fast $E_N$  can be allowed to
grow, in the case when the couplings are Gaussian.

\thanks  We would like to thank Stephan Mertens for interesting
 discussions.

\bigskip

\chap{2. Abstract theorems.}2

In this section we will  formulate a general result that implies the REM property
under some concise conditions, that can  be verified in
concrete examples. This will also allow us to present the broad outline
of the structure of the proof without having to bother with
technical details. Note that our approach is rather different from
that of \cite{BCMN} that involves computations of moments.

Our approach to the proof of the Mertens conjecture is based on
the following theorem, which
provides a criterion for Poisson convergence in a rather
general setting.

\theo{\TH(mainth)}  {\it Let $V_{i,M} \geq 0$,
    $i\in \N$, be a family of non-negative random
     variables satisfying the following assumptions:
for any $\ell\in \N $
 and all sets of  constants  $b_{j}>0$,  $j=1,\dots, \ell$,
$$
\lim_{M\uparrow\infty}\sum_{(i_1,\dots,i_\ell)\subset \{1,\dots,M\}}\P
  (\forall_{j=1}^\ell\,
V_{i_j, M} <b_j)\to \prod\limits_{j=1}^{\ell}b_{j}
\Eq(maincond)
$$
 where the sum is taken
   over all possible  sequences of {\sl different\/}
    indices
  $(i_1,\dots, i_\ell)$.
Then the point process
$$
\PP_M=\sum_{i=1}^M \delta_{V_{i,M}},
\Eq(pp)
$$
  on $\R_+$,
converges  weakly in distribution,  as $M\uparrow\infty$,
to the standard Poisson point process,
$\PP$ on $\R_+$ (i.e., the Poisson point process
   whose  intensity measure is the Lebesgue measure).
}

\remark Theorem \thv(mainth) was proven (in a more
    general form, involving vector valued random variables)
   in \cite{BK2}.
  It is  very similar in its spirit  to an
    analogous theorem for  the case of exchangeable variables
proven in \cite{BM} in an application to the Hopfield model.
The rather simple proof in the scalar setting can be found in Chapter
    13 of \cite{B}.

Naturally, we want to  apply this theorem with $V_{i,M}$ given by
$|N^{-1/2}H_N(\s)-E_N|$, properly normalised.

We will now introduce a setting in which the assumptions of Theorem
\thv(mainth) are verified.
Consider a product space $\SS^N$ where $\SS$ is a finite set.
 We define on $\SS^N$  a real-valued random   process, $Y_N(\s)$.
Assume for simplicity that
 $$
\E Y_N(\s)=0,\ \    \E(Y_N(\s))^2=1.
\Eq(ev)
$$
Define on $\SS^N$
$$
b_N(\s,\s')\equiv \cov(Y_N(\s),Y_N(\s')).
\Eq(ab.abs.1)
$$
Let us also introduce the Gaussian process,
  $Z_N$,  on $\SS^N$, that has the same  mean and  the same covariance matrix
  as $Y_N(\s)$.

Let $G$ be the group of automorphisms on $\SS_N$, such that, for
$g\in G$, $Y_N(g\s)=Y_N(\s)$, and let $F$ be the larger group, such that, for
$g\in F$,  $|Y_N(g\s)|=|Y_N(\s)|$.
Let
$$
E_N= c N^{\a},\ \ \ c,\a \in \R,\ \  0\leq \a<1/2,
\Eq(ab.abs.2)
$$
be a sequence of real numbers, that is either a constant,
 $c\in \R$,  if $\a=0$,   or
converges to plus or minus infinity, if $\a>0$.
 We will define  sets $\Sigma_N$ as follows:
If $c \ne 0$, we denote by
  $\S_N$ be the set of residual classes of
$\SS^N$ modulo $G$; if $c= 0$, we let
$\S_N$ be the set of residual classes modulo $F$.
We will assume throughout
  that $|\S_N|> \kappa^N$, for some $\kappa>1$.
   Define the sequence of numbers
 $$
\d_N= \sqrt{\sfrac {\pi}2} e^{ E_N^2/2} |\S_N|^{-1}.
\Eq(ab.abs.2.1)
$$
   Note that $\d_N$ is   exponentially small in  $N \uparrow \infty$,
   since $\a<1/2$. This sequence
   is chosen such that, for any $b\geq 0$,
$$
\lim_{N\uparrow\infty}|\S_N| \P(|Z_N(\s)-E_N|< b \d_N ) =b.
\Eq(ab.abs.2.2)
$$
  For $\ell\in\N$, and any collection, $\s^1,\dots,\s^\ell\in
  \S_N^{\otimes \ell}$,
we denote by $B_N(\s^1,\dots,\s^\ell)$ the covariance matrix
  of $Y_N(\s)$  with elements
$$
b_{i,j}(\s^1,\dots, \s^{\ell})\equiv b_N(\s^i,\s^j).
\Eq(ab.abs.41)
$$

\noindent {\bf Assumptions A.} {\it
\item{(i)}
Let $\RR_{N,\ell}^\eta$ denote the set
$$
\RR^\eta_{N,\ell}\equiv
 \{ (\s^1,\dots,\s^\ell)\in\S_N^{\otimes \ell}: \, \forall_{1\leq
i<j\leq \ell}\ |b_N(\s^i,\s^j)|\leq N^{-\eta} \}.
\Eq(ab.abs.6)
$$
Then  there exists a continuous decreasing
   function, $\rho(\eta)>0$,
 on $]\eta_0, \tilde \eta_0[$  (for some $
   \tilde \eta_0 \geq \eta_0>0$),  and
    $\mu>0$,  such that
$$
|\RR^\eta_{N,\ell}|\geq \left(1-\exp\left({- \mu(\eta) N^{\rho(\eta)}}
\right)\right) |\S_N|^\ell.
\Eq(ab.abs.4)
$$
\item{(ii)} Let $\ell \geq 2$,  $r=1,\dots, \ell-1$.
Let
 $$
\eqalign{
\LL^\ell_{N,r}= \Big\{(\s^1,\dots,\s^\ell)\in
   \S_N^{\otimes \ell} : \ & \forall_{1\leq i<j \leq \ell}\,
      |Y_N(\s^i)| \ne |Y_N(\s^j)|,  \cr
& \ \ \  \rank(B_N(\s^1,\dots, \s^{\ell}))=r
 \Big\}
}
\Eq(ab.abs.4.1)
$$
Then there exists $d_{r, \ell}>0$, such that,
  for all $N$ large enough,
$$
|\LL^\ell_{N,r}|\leq |\S_N|^r e^{-d_{r,\ell} N }.
\Eq(ab.abs.8)
$$
\item{(iii)}  For any $\ell\geq 1$, any  $r=1,2,\dots, \ell$,
    and any $b_1,\dots, b_{\ell}\geq 0$, there exist  constants,
$p_{r, \ell}\geq 0$  and
    $Q< \infty $, such that,
 for any $\s^1,\dots, \s^{\ell} \in \S_N^{\otimes \ell}$ for which
    $\rank(B_N(\s^1,\dots,\s^\ell)) =r$,
$$
\P \big (\forall_{i=1}^{\ell} :
   |Y_N(\s^i)-E_N |<\d_N b_i \big)
      \leq   Q \d_N^r  N^{p_{r, \ell}}.
\Eq(ab.abs.7)
$$
}

 \theo{\TH(ABS.1)}{\it Assume the Assumptions  A  hold.
  Assume that $\a \in [0, 1/2[$  is such that,
    for some $\eta_1\leq\eta_2 \in ]\eta_0, \tilde \eta_0[$,
  we have:
$$
\a <\eta_2/2,
\Eq(ab.abs.9)
$$

$$
\a <\eta/2+\rho(\eta)/2,  \,  \forall \eta \in ]\eta_1 , \eta_2[,
\Eq(ab.abs.10)
$$
and
$$
\a<\rho(\eta_1)/2.
\Eq(ab.abs.10a)
$$
       Furthermore,  assume that, for any  $\ell \geq 1$,
 any $b_1,\dots, b_{\ell}>0$,  and
    $(\s^1,\dots, \s^{\ell}) \in \RR^{\eta_1}_{N, \ell}$,
$$
\P \left(\forall_{i=1}^{ \ell}:\
    |Y_N(\s^i)-E_N |< \d_N b_i \right)=
\P \left(\forall_{i=1}^{ \ell }:\
   |Z_N(\s^i)-E_N |<  \d_N b_i \right)+o(|\Sigma_N|^{-\ell}).
\Eq(ab.abs.5)
$$
Then,  the point process,
$$
\PP_N\equiv \sum_{\s\in \S_N} \d_{\{ \delta_N^{-1}
|Y_N(\s)-E_N|\}} \rightarrow\PP 
\Eq(ab.abs.11)
$$
  converges weakly to the standard Poisson point process
 $\PP$ on $\R_+$.

\noindent Moreover, for any $\e>0$ and any $b\in \R_+$,
$$
\P \left(\forall_{N_0} \exists_{N\geq N_0} :\exists_{\s,\s':
|b_N(\s,\s')|>\e} : | Y_N(\s)-E_N|\leq |Y_N(\s')-E_N|\leq \d_N b \right) =0.
\Eq(ab.abs.12)
$$
}

\remark Before giving the proof of the theorem, let us comment on the
various assumptions.
\item{(i)} Assumption A (i) holds with some $\eta$ in any reasonable
model, but the function $\rho(\eta)$ is model dependent.
\item{(ii)} Assumptions A (ii) and (iii) is also apparently valid in
most cases, but can be tricky sometimes. An example where (ii)  proved
difficult is the $k$-partitioning problem, with $k>2$.
\item{(iii)} Condition \eqv(ab.abs.12)
is essentially a local central limit theorem. In the case $\a=0$
it  holds, if the Hamiltonian is a sum over independent random
interactions,  under mild decay assumptions on the characteristic
function of the distributions of the interactions.
 Note that some such assumptions are obviously necessary,
  since if the random interactions take on only
finitely many values, then also the Hamiltonian will take values
on a lattice, whose spacings are not exponentially small, as would
be necessary for the theorem to hold.
 Otherwise, if $\a>0$, this will require further assumptions on the interactions.
We will leave this problem open in the present paper.
 It is of course trivially verified, if the interactions are
Gaussian.

\proof We just have to verify the hypothesis of Theorem \thv(mainth),
for $V_{i,M}$ given by $\delta_N^{-1}|Y_{N}(\s)-E_N|$, i.e., we must show that
$$
\sum_{(\s^1,\dots,\s^\ell) \in \S_N^{\otimes l}}
\P \left(\forall_{i=1}^{\ell}:
 |Y_N(\s^i)-E_N|< b_i \d_N \right) \rightarrow b_1\cdots b_\ell.
\Eq(ab.abs.13)
$$
We  split this sum into the sums over
  the set $\RR^{\eta_1}_{N,\ell}$ and its complement.
First, by the assumption \eqv(ab.abs.5)
$$
\eqalign{
& \sum_{(\s^1,\dots,\s^\ell)\in \RR^{\eta_1}_{N,\ell}}
\P \left(\forall_{i=1}^{ \ell} : \
|Y_N(\s^i)-E_N |< b_i \d_N\right) \cr
& =
\sum_{(\s^1,\dots,\s^\ell)\in \RR^{\eta_1}_{N,\ell}}
\P \left(\forall_{i=1}^{ \ell }:\
|Z_N(\s^i)-E_N |< b_i \d_N\right)+o(1).}
\Eq(ab.abs.14)
$$
 But, with
   $\CC(E_N)=\{\vec x=(x_1,\dots, x_\ell)\in \R^{\ell}:\forall_{i+1}^\ell
 |E_N-x_i|\leq  \d_N b_i \}$,
$$
\P\left(\forall_{i=1}^{\ell} :
   |Z_N(\s^i)-E_N |< b_i \d_N \right)
=\int\limits_{\CC(E_N)}
   \frac{e^{-(\vec z,B_N^{-1}(\s^1,\dots,\s^\ell) \vec z)/2 }}
{(2\pi)^{\ell/2}\sqrt{\det(B_N(\s^1,\dots,\s^\ell))}}d\vec z,
\Eq(ab.abs.15)
$$
where $B_N(\s^1,\dots,\s^\ell)$ is the covariance matrix defined in
\eqv(ab.abs.41).
Since $\d_N $ is  exponentially small in $N$,
we see that, uniformly for $(\s^1,\dots,\s^{\ell})
  \in \RR^{\eta_1}_{N, \ell}$,
 the integral \eqv(ab.abs.15) equals
$$
  (2 \delta_N /\sqrt{2\pi})^{\ell} (b_1 \cdots b_{\ell})
e^{- (\vec E_N,B^{-1}(\s^1,\dots,\s^\ell) \vec E_N)/2 }(1+o(1)),
\Eq(abs.n)$$
  where we denote by $\vec E_N$ the vector
  $(E_N, \dots, E_N)$.

  We treat separately
the  sum \eqv(ab.abs.14) taken
  over the smaller set,  $\RR^{\eta_2}_{N, \ell} \subset
   \RR^{\eta_1}_{N, \ell}$,
  and the one over
$\RR^{\eta_1}_{N, \ell}\setminus \RR^{\eta_2}_{N, \ell}$.

  Since,  by \eqv(ab.abs.9), $\eta_2$ is chosen
such that
$ E_N^2 N^{-\eta_2}\to 0  $,
   by \eqv(ab.abs.5), \eqv(ab.abs.15), and \eqv(abs.n),
  each term in the sum over $\RR^{\eta_2}_{N, \ell}$
  equals
$$
(2 \delta_N/\sqrt{2\pi})^{\ell} (b_1 \cdots b_{\ell}) e^{-\frac{1}{2}
  \|\E_N\|^2 (1+O(N^{-\eta_2}))}(1+o(1))= (b_1 \cdots
  b_{\ell})|\Sigma_N|^{-\ell}(1+o(1)),
\Eq(ab.heur.3)
$$
  uniformly for  $(\s^1, \dots, \s^{\ell}) \in \RR^{\eta_2}_{N, l}$.
 Hence by  Assumption A (i)
$$
\eqalign{
 \sum_{(\s^1,\dots,\s^\ell)\in \RR^{\eta_2}_{N,\ell}}
\P \left(\forall_{i=1}^{\ell} :
|Z_N(\s^i)-E_N |< b_i \d_N\right) &=
|\RR^{\eta_2}_{N, l}|  |\Sigma_N|^{-\ell}
   (b_1\cdots b_{\ell})(1+o(1))\cr
&\to b_1 \cdots b_l.
}
\Eq(yuy)
$$
   Now let us consider the remaining  set
 $\RR^{\eta_1}_{N, \ell}\setminus \RR^{\eta_2}_{N, \ell}$
 (if it is non-empty, i.e., if strictly $\eta_1<\eta_2$),
    and let us
 find $\eta_1=\eta^{0}<\eta^{1}<\cdots <\eta^{n}=\eta_2$,
 such that
$$
\a < \eta^i/2+\rho(\eta^{i+1})/2 \ \ \  \forall i=0,1,\dots, n-1,
\Eq(sss)
$$
 which  is  possible due to the assumption \eqv(ab.abs.10).
      Then let us split the sum over
 $\RR^{\eta_1}_{N, l}\setminus \RR^{\eta_2}_{N,\ell}$
  into $n$ sums, each  over
    $\RR^{\eta_{i}}_{N, \ell}\setminus \RR^{\eta_{i+1}}_{N, \ell}$,
   $i=0,1, \dots,n-1$.
By \eqv(ab.abs.5), \eqv(ab.abs.15), and \eqv(abs.n), we have,
  uniformly for $(\s^1,\dots, \s^{\ell}) \in \RR^{\eta^{i}}_{N, \ell}$,
$$
 \eqalign{ \P \left(\forall_{i=1}^{\ell} :
|Z_N(\s^i)-E_N |< b_i \d_N\right)
&  =
    (2 \delta_N/\sqrt{2\pi})^{\ell} (b_1 \cdots b_{\ell}) e^{-\frac{1}{2}
  \|\E_N\|^2 (1+O(N^{-\eta^i}))}(1+o(1))\cr
& \leq C |\S_N|^{-\ell} e^{N^{2\a -\eta^i}},
}
\Eq(ab.abs.35)
$$
for some constant $C<\infty$.
 Thus by  Assumption A (i),
$$
\eqalign{
 \sum_{ \RR^{\eta_{i}}_{N, l}\setminus \RR^{\eta_{i+1}}_{N,l} }
     \P (\forall_{i=1}^{\ell} :
|Z_N(\s^i)-E_N |< b_i \d_N )  & \leq  C
     |\S_N^{\otimes l}\setminus \RR^{\eta_{i+1}}_{N,l}|
  |\S_N|^{-\ell} e^{N^{2\a -\eta^i}} \cr
  &  \leq C \exp\left({- \mu(\eta^{i+1}) N^{\rho(\eta^{i+1})} }+N^{2\a
 -\eta^i}
\right),
}
\Eq(ab.abs.35.1)
$$
  that, by \eqv(sss), converges to zero,
 as $N \to \infty$,  for any $i=0,1,\dots, n-1$.
  So the sum \eqv(ab.abs.14) over
   $\RR^{\eta_1}_{N, l}\setminus \RR^{\eta_2}_{N,l}$
  vanishes.

      Now we turn to the sum over collections,
  $(\s^1,\dots, \s^{\ell}) \not \in \RR_{N, l}^{\eta_1}$.
  We
distinguish the cases when $\det(B_N(\s^1,\dots,\s^\ell))=0$ and
 $\det(B_N(\s^1,\dots,\s^\ell))\neq 0$.
For the contributions from the latter case, using Assumptions A (i) and (iii),
  we  get readily that,
$$
\eqalign{
\sum_{{(\s^1,\dots,\s^\ell)\not\in \RR^{\eta_1}_{N\ell}}
   \atop {\hbox{\sm rank} (B_N(\s^1,\dots, \s^{\ell}))=\ell}}
&  \P \left(\forall_{i=1}^\ell\,
|Y_N(\s^i)-E_N |<\d_N b_i \right)
 \leq |\S_N|^\ell e^{-\mu(\eta_{1}) N^{\rho(\eta_1)}}
Q |\d_N |^{\ell} N^{p_\ell}  \cr
& \leq C  N^{p_{\ell}}\exp\left({-\mu(\eta_{1}) N^{\rho(\eta_1)}
  +\ell E_N^2/2}\right).
}
\Eq(ab.abs.17)
$$
The right-hand side of \eqv(ab.abs.17) tends to zero
  exponentially fast,  if condition
\eqv(ab.abs.10a) is verified.

Finally, we must deal with the contributions from the cases when the
covariance matrix is degenerate, namely
$$
\sum_{ {(\s^1,\dots, \s^{\ell})   \in \S_N^{\otimes l}} \atop
 {\hbox{\sm rank}(B_N(\s^1,\dots, \s^{\ell}))=r }}
 \P (\forall_{i=1}^\ell :
|Y_N(\s^i)-E_N |< b_i \d_N),
\Eq(sssss)
$$
for $r=1,\ldots, \ell-1$.
 In the case $c=0$,  this sum is taken
    over the set $\LL^r_{N, \ell}$,
 since   $\s$ and $\s'$ in $\S_N$ are different,
  iff $|Y_N(\s)| \ne |Y_N(\s')|$, by definition of $\S_N$.
  In the case  $c\ne 0$,
  this sum is taken  over  $\ell$-tuples
   $(\s^1,\dots, \s^{\ell})$ of different elements
   of $\Sigma_N$, i.e., such that $Y_N(\s^i) \ne Y_N(\s^j)$,  for any
  $1\leq i< j \leq \ell $.
  But for all $N$ large enough,
       all terms in this sum over  $\ell$-tuples,
   $(\s^1, \dots, \s^{\ell})$,
  such that  $Y_N(\s^i) =- Y_N(\s^j)$,
  for some   $1\leq i< j \leq \ell $,
   equal zero, since
   the events $\{|Y_N(\s^i)- E_N|<b_i \d_N\}$
  and  $\{|-Y_N(\s^i)- E_N|<b_j \d_N\}$, with $E_N=c N^{\a}$,
   $c\ne 0$, are disjoint.
  Therefore \eqv(sssss) is reduced to the sum over
  $\LL^r_{N, \ell}$ in the case $c\ne 0$
    as well. Then, by  Assumptions A (ii)
  and (iii), it is bounded from above by
$$
 |\LL^r_{N, \ell}| Q (\d_N )^{r} N^{p_{r, \ell}}
 \leq  |\S_N|^{r} e^{-d_{r,\ell}N} Q (\d_N)^{r}
  N^{p_{r, \ell}}
\leq
C e^{-d_{r,\ell}N}
e^{\ell E_N^2/2} N^{p_{r, \ell}}.
\Eq(ab.abs.18)
$$
  This bound  converges to zero
  exponentially fast,   since $E_N^2= c^2 N^{2\a}$, with $\a<1/2$.
This concludes the proof of the first part of the theorem.

The second assertion \eqv(ab.abs.12)
 is elementary: by \eqv(ab.abs.17) and \eqv(ab.abs.18),
 the sum of terms $\P(\forall_{i=1}^2 : |Y_N(\s^i)-E_N| <\delta_N b)$
  over all pairs,
    $(\s^1, \s^2) \in \Sigma_{N}^{\otimes 2}
  \setminus \RR_{N, 2}^{\eta_1}$, such that  $\s^1 \ne \s^2$,
 converges to zero exponentially fast. Thus \eqv(ab.abs.12)
    follows from the Borel-Cantelli lemma.
\endproof

  Finally, we  remark that the
  results of Theorem  \thv(ABS.1) can be extended
to the case when  $\E Y_N(\s)\ne 0$, if $\a=0$, i.e., $E_N=c$.
Note that, e.g. the unrestricted number partitioning problem falls
  into this class.
      Let now $Z_N(\s)$ be the Gaussian process
  with the same mean and covariances as $Y_N(\s)$.
Let us consider both the covariance matrix, $B_N$, and the mean of
  $Y_N$,
 $\E Y_N(\s)$, as random variables on the probability space
$(\S_N,\BB_N,\E_\s)$, where $\E_\s$ is the uniform law
on $\S_N$.
 Assume that, for any $\ell\geq 1$,
$$
B_N(\s^1,\dots, \s^{\ell}) \limlaw I_d, \ \ N \uparrow \infty,
   \Eq(ab.z1)
$$
   where $I_d$ denotes the identity matrix,  and
$$
\E Y_N(\s) \limlaw D, \ \ N \uparrow \infty,
 \Eq(ab.z2)
$$
 where $D$ is some random variable $D$.
Let
$$ \widetilde \delta_N =\sqrt{\sfrac \pi2}K^{-1}
    |\Sigma_N|^{-1}.
\Eq(vbvb)
$$
 where
$$
K\equiv \E e^{-(c-D)^2/2}.
\Eq(ab.z3)
$$

\theo{\TH(ABS.3)}{\it  Assume that, for some
  $R>0$, $|\E Y_N(\s)| \leq N^R$, for all $\s \in \S_N$.
Assume that \eqv(ab.abs.4)  holds for some $\eta>0$ and
  that (ii) and (iii) of Assumptions A are valid.
Assume that there exists a set, $\QQ_{N} \subset \RR^{\eta}_{N,
\ell}$,
 such that
 \eqv(ab.abs.5) is valid   for any
   $(\s^1, \dots, \s^{\ell})\in \QQ_{N}$, and that
$|\RR_{N,\ell}^{\eta} \setminus \QQ_{N}|
   \leq |\Sigma_N|^{\ell} e^{-N^{\gamma}}$,
   with some $\gamma>0$.
 Then, the point process
$$
\PP_N\equiv \sum_{\s\in \S_N} \d_{ \widetilde
   \delta_N^{-1} |Y_N(\s)-E_N|}\rightarrow\PP
\Eq(ab.abs.11.1)
$$
    converges weakly to the standard Poisson point process
 $\PP$ on $\R_+$ . }

\noindent{\it Proof.} We must prove again the convergence of the sum
\eqv(ab.abs.13), that we split into three sums:  the first over
  $\QQ_{N}$, the second over  $\RR^{\eta}_{N, \ell} \setminus \QQ_{N}$,
  and the third over the complement of  the set   $\RR^{\eta}_{N, \ell}$.
  By  assumption, \eqv(ab.abs.5) is valid on $\QQ_{N}$,
   and thus  the terms of the first sum  are reduced to
$$
\eqalign{
\int\limits_{\forall i=1,\dots, \ell:
     |z_i- c|< \widetilde \delta_N b_i  } &
   \frac{e^{-((\vec z- \E \vec Y_N(\s)) B_N^{-1}(\s^1,\dots,\s^\ell)
   (\vec z- \E \vec Y_N(\s)))/2 }}
{(2\pi)^{\ell/2}\sqrt{\det(B_N(\s^1,\dots,\s^\ell))}}d\vec z
  \cr &
\kern -1cm= (2 \tilde \delta_N/\sqrt{2\pi} )^{\ell} (b_1 \cdots
b_{\ell})
  e^{-( \vec c- \vec \E Y_N(\s))  B^{-1}(\s^1,\dots,\s^\ell)
  (\vec c-\E \vec Y_N(\s))/2}(1+o(1)),
 }
\Eq(kjk)
$$
 with $\vec c\equiv(c,\dots, c)$, and
   $E \vec Y_N(\s)\equiv (\E Y_N(\s^1),\dots, \E Y_N(\s^{\ell}))$,
  since $\delta_N$ is exponentially small and  $|\E Y_N(\s)| \leq N^R$.
By definition of $\tilde \delta_N$, the quantities \eqv(kjk)
  are at most $O(|\S_N|^{-\ell})$, while, by the estimate
   \eqv(ab.abs.4) and by the assumption on the cardinality of
     $\RR^{\eta}_{N, \ell} \setminus \QQ_{N}$,  the number of
   $\ell$-tuples of
    configurations in $\Sigma_N^{\otimes l} \setminus \RR^{\eta}_{N, \ell}$
  and in  $\RR^{\eta}_{N, \ell} \setminus \QQ_{N}$
 is exponentially smaller than $|\Sigma_N|^{\ell}$.
     Hence
$$
\eqalign{
& \sum_{(\s^1,\dots,\s^\ell)\in \QQ_{N}}
\P (\forall_{i=1}^{\ell} :
|Y_N(\s^i)-E_N |< b_i \d_N) \cr
& = \sum_{(\s^1,\dots,\s^\ell)\in \QQ_{N}}
(2 \tilde \delta_N/\sqrt{2\pi} )^{\ell} (b_1 \cdots b_{\ell})
e^{-( \vec c- \vec \E Y_N(\s))  B^{-1}(\s^1,\dots,\s^\ell)
 (\vec c-\E \vec Y_N(\s))/2
} (1+o(1))+o(1)\cr
&= \sum_{(\s^1,\dots,\s^\ell) \in \Sigma_N^{\otimes \ell} }
(2 \tilde \delta_N/\sqrt{2\pi} )^{\ell} (b_1 \cdots b_{\ell})
e^{-( \vec c- \vec \E Y_N(\s))  B^{-1}(\s^1,\dots,\s^\ell)
 (\vec c-\E \vec Y_N(\s))/2
} (1+o(1))+o(1)\cr
&= \frac{b_1\cdots b_\ell} {|\S_N|^{\ell}K^{\ell}}
   \sum_{(\s^1,\dots,\s^\ell) \in \Sigma_N^{\otimes \ell} }
 e^{-( \vec c- \vec \E Y_N(\s))  B^{-1}(\s^1,\dots,\s^\ell)
 (\vec c-\E \vec Y_N(\s))/2
} (1+o(1))+o(1).
 }
\Eq(ab.abs.mean.1)
$$
 The last quantity converges  to $b_1\cdots b_{\ell}$,
    by the assumptions
  \eqv(ab.z1), \eqv(ab.z2) and \eqv(ab.z3).

    The  sum of the probabilities,
   $\P(\forall_{i=1}^{\ell}: |Y_N(\s)-E_N|<\delta_N b_i)$,
   over all $\ell$-tuples of  $\RR^{\eta}_{N, \ell} \setminus \QQ_{N}$,
   contains at most $|\Sigma_N|^{\ell} e^{-N^{-\gamma}}$
   terms, while, by Assumption A (iii),
  (and since, for any $(\s^1,\dots, \s^{\ell}) \in \RR^{\eta}_{N, \ell}$,
  the rank of $B_N(\s^1,\dots, \s^{\ell})$ equals $\ell$)
   each term
   is at most of  order $|\Sigma_N|^{-\ell}$, up to a polynomial factor.
    Thus this sum converges to zero.

     Finally, the sum of the same probabilities over the collections
$(\s^1,\dots,\s^\ell) \in \Sigma_N^{\otimes l}\setminus
     \RR^{\eta }_{N,\ell}$
  converges to zero, exponentially fast,
  by the same arguments as those leading to \eqv(ab.abs.17) and \eqv(ab.abs.18),
  with $\eta_1=\eta$.
 \endproof

\bigskip

\chap{3. Examples}3

We will now show that the assumptions of our theorem are verified in a
wide class of physically relevant models: 1) the Gaussian $p$-spin SK
models, 2) SK-models with non-Gaussian couplings,
and 3) short-range spin-glasses. In the last two examples we consider
only  the case $\a=0$.

\noindent{\bf 3.1 $p$-spin Sherrington-Kirkpatrick models,
    $0\leq \a<1/2$.}

 In this subsection we illustrate  our general theorem in
 the class of Sherrington-Kirkpatrick models.
  Consider $\SS=\{-1, 1\}$.
$$
  H_N(\s)=\frac{\sqrt{N}}{ \sqrt{ N \choose p }}\sum_{1\leq
i_1<i_2<\dots <i_p
 \leq N }
   J_{i_1,\dots, i_p} \s_{i_1}\s_{i_2}\cdots \s_{i_p}
\Eq(ab.tt1.3)
$$
  is the Hamiltonian of the $p$-spin Sherrington-Kirkpatrick model,
 where $J_{i_1,\dots, i_p}$ are independent
    standard Gaussian random variables.

 The following elementary proposition concerns the symmetries to
the Hamiltonian.

\proposition{\TH(pr)}{\it
  Assume that, for any $0<i_1<\dots <i_p\leq N$,
     $\s_{i_1}\cdots \s_{i_p} =\s'_{i_1}\cdots \s'_{i_p}$.
  Then, if $p$ is pair,
   either $\s_i=\s_i'$, for all $i=1,\dots, N$, or $\s_i=-\s'_i$,
   for all $i=1,\dots, N$, and, if $p$ is odd, then
     $\s_i=\s_i'$, for all $i=1,\dots, N$.
   Assume that, for any $0<i_1<\dots <i_p\leq N$,
      $\s_{i_1}\cdots \s_{i_p} = -\s'_{i_1}\cdots \s'_{i_p}$.
 This is impossible, if $p$ is pair  and
    $\s_i=-\s_i'$, for all $i=1,\dots, N$, if
    $p$ is odd. }

  This proposition allows us to construct
   the space $\Sigma_N$:
 If $p$ is odd and $c \ne 0$,
  $\Sigma_N=\SS^N$, thus  $|\Sigma_N|=2^N$.
         If $p$ is even,  or $c=0$,
      $\Sigma_N$ consists of equivalence classes
   where configurations
    $\s$ and $-\s$ are identified, thus
      $|\Sigma_N|=2^{N-1}$.

\theo{\TH(unp-not0)}{\it  Let $p\geq 1$ be odd.
    Let $\Sigma_N =\SS^N$.
          If $p=1$ and
   $\a \in [0, 1/4[$, and, if $p=3,5,\dots,$,
   and  $\a \in [0, 1/2[$, for any constant $c \in \R\setminus\{0\}$
   the point process
$$
\PP_N\equiv
\sum_{\s \in \Sigma_N } \delta_{\{ \d_N^{-1}
   |N^{-1/2} H_N(\s) - c N^{\a } | \}}
\Eq(pp1)
  $$
where
$\d_N=2^{-N} e^{+c^{2} N^{2\a }/2}
 \sqrt{\sfrac\pi 2}$,
converges weakly to the standard Poisson point process, $\PP$,  on $\R_+$.

Let $p$ be even. Let $\Sigma_N$ be the space of equivalence classes
   of  $\SS^N$ where $\s$ and $-\s$ are identified.
        For any
      $\a \in [0, 1/2[$ and any constant, $c\in \R$,
   the point process
$$
\PP_N\equiv\sum_{\s \in \Sigma_N } \delta_{\{(2\d_N)^{-1}| N^{-1/2}
   H_N(\s)- c N^{\a} | \}}
\Eq(pp2n)
$$
 converges weakly to the standard Poisson point process, $\PP$, on $\R_+$.
  The result \eqv(pp2n) holds true as well  in the
  case of $c=0$, for  $p$ odd.
}

%\end

\noindent {\it Proof of Theorem} \thv(unp-not0).
      We have to verify the assumptions of Theorem~\thv(ABS.1)
 for the process $N^{-1/2}H_N(\s)=Y_N(\s)$.
     The elements of the covariance matrix \eqv(ab.abs.41) are:
 $$
b_{j,j}(\s^1,\dots,\s^{\ell} ) =1,\ \ \forall_{j=1}^{\ell };
\Eq(ii)
$$
$$
b_{j,m}(\s^1,\dots, \s^{\ell}) = {N \choose p}^{-1} \sum_{1\leq i_1<i_2
 <\cdots<i_p\leq N }\s_{i_1}^j\dots \s_{i_p}^j
    \s_{i_1}^m\dots \s_{i_p}^m, \  \forall_{1\leq j<m \leq \ell}.
\Eq(tt)
 $$
   It has been observed in \cite{BKL} that
its non-diagonal elements can be written as
$$
b_{j,m}(\s^1,\dots, \s^{\ell })= \sum_{k=0}^{[p/2]} (-N)^{-k}
{2k \choose p} (k-1)!!
  \Big(\sfrac 1N{\sum_{q=1}^N \s^j_q \s^m_q}\Big)^{p-2k}(1+O(1/N)).
\Eq(ij)
$$

      Now let us verify the Assumption A (i). Let
$$
\QQ^{\zeta}_{N, \ell, q}=\Big\{(\s^1,\dots, \s^l) \in \S_N^{\otimes l} :
  \forall_{ 1\leq i<j \leq \ell} \
    \Big| N^{-1}\sum_{q=1}^{N} \s_q^i \s_q^j \Big |< q
  N^{-\zeta}\Big\}.
\Eq(qqqqq)
$$
  The $\ell$-tuples of this set satisfy the following
     property:
   for any $\delta_2,\dots, \delta_\ell \in \{-1,1\}^{\ell-1}$,
    the sets of sites
          $A_{\delta_2,\dots, \delta_\ell} =
   \{i: \s_i^2=\delta_2 \s_i^1, \s_i^3=\delta_3 \s_i^1,
\dots, \s_i^\ell=\delta_\ell \s_i^1 \}$
   has the cardinality $N 2^{-(\ell-1)}+ O(N^{1-\zeta})$.
      Then it is an easy combinatorial
  computation to check that there exists $h>0$, such that,
 for any  $q \in \R_+$,
and any $\zeta \in ]0, 1/2[$,
$$
|\QQ^{\zeta}_{N, \ell,q}| \geq |\S_N|^{\ell} (1-\exp(-h
N^{1-2\zeta})),
\Eq(aaaa)
$$
 for all $N$ large enough.
   By the representation \eqv(ij), we have
  $\bigcap_{k=0}^p \QQ^{-(k-\eta)/(p-2 k)}_{N, \ell, q}
 \subset  \RR^{\eta}_{N, \ell} $,  with
   $q=(p \max_{k=0, \dots, [p/2]}{2k \choose p}(k-1)!!)^{-1}$.
  But, for any $\eta \in ]0, p/2[$, and any $k=0,1,\dots, [p/2]$,
    $ \QQ^{-(k-\eta)/(p-2 k)}_{N, \ell,q} \subset
     \QQ^{-(k+1-\eta)/(p-2 (k+1))}_{N, \ell,q}$.
   Therefore,
  $$
\QQ^{\eta/p}_{N, \ell, q} \subset
    \RR^{\eta}_{N, \ell}.
\Eq(qqqq)
$$
Thus, due to \eqv(aaaa),
   Assumption  A (i) is verified with  $\rho(\eta)= 1-2 \eta/p$,
  for  $\eta \in ]0, p/2[$.

     Let us now check the Assumption A (ii).
  To estimate the cardinality of  $\LL^r_{N, \ell}$,
      we need to introduce an $\ell$ by ${N \choose p}$  matrix,
  $C_p(\s^1,\dots, \s^{\ell})$, as follows.
   For any given $\s^1,\dots, \s^{\ell}$,
 the $j$th column is  composed of all ${N \choose p}$
 products,  $\s_{i_1}^j\s_{i_2}^j \cdots \s_{i_p}^j$,
  over all subsets $1\leq i_1<i_2<\cdots <i_p \leq N$.
   Then we have
$$
C^T_p(\s^1,\dots, \s^{\ell}) C_p(\s^1,\dots, \s^{\ell})
  = {N \choose p}^{-1} B_N(\s^1,\dots, \s^{\ell}).
\Eq(ccb)
$$
      Let $\s^1,\dots, \s^{\ell}$ be such that
  $\rank(B_N(\s^1,\dots, \s^{\ell}))=r<\ell$.
       Then, $r$ columns
  of the matrix $C_p(\s^1,\dots, \s^{\ell})$
  form a basis of its $\ell$ columns.
      Assume that these are, e.g., the first $r$ columns.
         The matrix $C_p(\s^1,\dots, \s^r)$
  can contain at most $2^r$ different rows.
         We will show that, for any
  $(\s^1,\dots, \s^{\ell}) \in \LL^r_{N, \ell}$,
       it can in fact not contain all
  $2^r$ rows, due to the following  proposition.

\proposition{\TH(prs)} {\it Assume that
  an $2^r \times r$  matrix, $A$, with
  elements, $1$ or $-1$,  consists
    of all $2^r$ different rows.
          Assume that a column of length $2^r$
  with elements $1$ or $-1$  is
  a linear combination of the columns of $A$.
     Then this column is a multiple (with coefficient $+1$ or $-1$) of
  one of the
  columns of the matrix $A$. }

\noindent{\it Proof.}  The proof can be carried
  out by induction over $r$. A generalisation
  of this fact is proven in \cite{BK-npp}.

       Now,  if the matrix $C_p(\s^1,\dots, \s^r)$ contained all
  $2^r$ rows, then, by Proposition~\thv(prs),
     for any $j=r+1,\dots, \ell$, there would exist
   $m=1,\dots, r$, such that,  either,
   for any $0<i_1<\dots <i_p\leq N$,
    $\s_{i_1}^j\cdots \s_{i_p}^j =\s_{i_1}^m
   \cdots \s_{i_p}^m$,  or, for any $0<i_1<\dots <i_p\leq N$,
     $\s_{i_1}^j\cdots \s_{i_p}^j = - \s_{i_1}^m
   \cdots \s_{i_p}^m$, which would imply
     $|Y_N(\s^j)|=|Y_N(\s^m)|$.
    But this is excluded by the  definition of $\LL^{r}_{N, \ell}$.

       Thus, for any $(\s^1,\dots, \s^{\ell}) \in \LL^{r}_{N, \ell}$,
   the matrix  $C_p(\s^1,\dots, \s^r)$
  contains at most $2^r-1$ different rows.
  There are $O((2^r-1)^N)$ possibilities
  to construct such a matrix.
      Furthermore, there is only an $N$-independent number of
   possibilities
  to complete it by adding linear combinations
  of its columns  to $C_p(\s^1,\dots, \s^{\ell})$.
        To see  this, consider the restriction
  of $C_p(\s^1,\dots, \s^r)$ to any $r$
  linearly independent
   rows. There are  not more than $2^{r(\ell-r)}$
        ways to complete it by $(\ell-r)$ columns of
   $\pm 1$ of length $r$, that are linear combinations
  of its $r$ columns.  But each such choice determines
  uniquely linear coefficients in these
  linear combinations and hence the completion
  of the whole $C_p(\s^1,\dots, \s^r)$ up to
  $C_p(\s^1,\dots, \s^{\ell})$.
    Thus  $|\LL^{r}_{N, \ell}| = O((2^r-1)^N)$.

  It remains to verify the Assumption A (iii).
   This is easy: if $\rank(B_N(\s^1,\dots, \s^{\ell}))=r$,
   then $r$ of the random variables  $Y_N(\s^1), \dots, Y_N(\s^{\ell})$
  are linearly independent. Assume that these are, e.g.,
    $Y_N(\s^{i_1}), \dots, Y_N(\s^{i_r})$. Then the covariance matrix
 $B_N(\s^{i_1},\dots, \s^{i_r})$ is non-degenerate, and
 the corresponding probability is bounded from above by
$$
\P(\forall_{j=1}^{r} |Y_N(\s^{i_j})- E_N|< \delta_N b_{i_j} ) \leq
    \frac{( 2 \delta_N)^r (b_{i_1}\cdots b_{i_r})}{
    \sqrt{(2\pi)^r \hbox{det}\,B_N(\s^{i_1},\dots, \s^{i_r}) }}.
\Eq(ab.sk.1)
$$
>From the representation of the matrix elements of $,B_N(\s^{i_1},\dots,
  \s^{i_r}))$, \eqv(tt),
one sees that  the determinant, $(\hbox{det}\,B_N(\s^{i_1},\dots,
  \s^{i_r}))$,
is a finite polynomial  in the variables $N^{-1}$, and thus
  its inverse
  can grow at most polynomially.

   Thus, we have established that  Assumption A is verified. We now
 turn to
conditions  \eqv(ab.abs.9), \eqv(ab.abs.10), and
  \eqv(ab.abs.10a) on $\a$.
     Since $\rho(\eta)=1-2 \eta/p$,
  for $\eta \in ]0, p/2[$, we should find
    $\eta_1, \eta_2 \in ]0, p/2[$ such that
 $\a<\eta_2/2$, $\a<\eta/2+1/2-\eta/p$ for
   $\eta \in ]\eta_1, \eta_2[$, and
    $\a<1/2-\eta_1/p$.
  We see that, for any $p\geq 2$ and $\a \in ]0, 1/2[$,
   it is possible to fix $\eta_1>0$ small enough,
     and $\eta_2 \in ]0, p/2[$ close enough to $1$,
  such that these assumptions are satisfied.
   If $p=1$,  then such a choice is possible only for  $\a \in ]0, 1/4[$.
   The assumption \eqv(ab.abs.5) need not be
   verified here  as $Y_N(\s)$ is a Gaussian process.
\endproof

\remark {\bf Values $p=1, \a=1/4$}.
The value $\a=1/4$ is likely to be the true critical value in the case
             $p=1$.
In this case, one can check that the principle part of our sum gives a
             contribution
of the form
 $$
\frac{ \hbox{const}(1+o(1))}{
\sqrt{(2\pi N)^{k(k-1)/2} } }
  \sum_{m_{1,2}, \dots, m_{k-1,k} \atop
  \forall i\ne j:
          |m_{i,j}|<N^{\eta-1/2} }
 \!\!\!\!\! \exp\Big( c^2 N^{2\a}
       \sum_{1\leq i<j \leq N}  m_{i,j}(1+o(1))
    - \frac{N}{2} \sum_{1\leq i<j \leq k} m_{i,j}^2(1+o(1))\Big),
 \Eq(ttn)
 $$
  which in turn is easily seem to be of order $(e^{c^2/2})^{k(k-1)/2
  }$,
that it it does not behave like a constant to the power $k$. Note that
  the term proportional to $\sqrt N$ in the exponents arises from the
off-diagonal part of the covariance matrix $B_N$.

      If $\a>1/4$, the contribution from the
   \eqv(ttn) is already of order $(e^{N^{4\a-1}c^2/2})^{k(k-1)/2
  }$, which cannot be compensated by any normalisation of the form
   $\d_N^k$. Thus at least the conditions of Theorem \thv(mainth)
   cannot hold in this case.

\bigskip

\noindent{\bf 3.2. Generalized $p$-spin SK models at level $\a=0$.}

\bigskip
     In this subsection we generalize
 Theorem~\thv(unp-not0) to the case
  of non-Gaussian process in the case of non-zero
  mean and $\a=0$.
     Let $p\geq 1$, $U_{i_1,i_2,\dots, i_p }$
       be any ${N \choose p}$ i.i.d. random variables
   with $\E U = a$ and $\hbox{Var}\,U =1$. Let
$$
H_N(\s)=\frac{\sqrt{N}}{ \sqrt{ N \choose p }}
    \sum_{1\leq i_1<i_2<\dots <i_p\leq N }
   U_{i_1,\dots, i_p} \s_{i_1}\s_{i_2}\cdots \s_{i_p}.
\Eq(ab.1)
$$
Let $\phi(s)=\E e^{i s (U-a)}$
   be the generating function of
   $(U-a)$.

\noindent{\bf Assumption B.}
 {\it   We will assume in this section
  that $E|U|^3<\infty$ and  $|\phi(s)|=O(|s|^{-1})$,
    as $|s|\to  \infty$. }

\remark  The decay assumption on the Fourier transform is not optimal,
  but some condition of this type is needed, as the result cannot be expected
  for discrete distributions, where the number of possible values the
  Hamiltonian takes on would be finite.

  We consider $Y_N(\s)= N^{-1/2}H_N(\s)$.
 The state space $\Sigma_N$ is defined as in the previous example.
  The covariance matrix, given by \eqv(ij),
  converges in law to the identity matrix by the law
 of large numbers. Furthermore, analogously to \eqv(ij),
   we see that $\E Y_N(\s)=Q_p( N^{-1/2} \sum_{i=1}^N \s_i)$,
   where
$$
Q_p(x)=  \sum_{k=0}^{[p/2]} (-1)^k
{2k \choose p} (k-1)!! x^{p-2k}.
\Eq(nij)
$$
   By the central limit theorem,
   $\E Y_N(\s) \limlaw Q_p(J)$
   where  $J$ is a standard Gaussian random variable.
   Hence, \eqv(ab.z1) and \eqv(ab.z2) are verified and we may
  define the constant
$$
K_p \equiv \E \exp\Big(-(c - a Q_p(J))^2/2 \Big)
\Eq(kp)
$$
Then, $\widetilde \delta_N = K_p^{-1}|\Sigma_N|^{-1} (\sqrt{2\pi}/2)$,
  with $|\Sigma_N|=2^N$ for $p$ odd and  $|\Sigma_N|=2^{N-1}$
  for $p$ even.

\theo{\TH(uthyyy)}{\it \item{(i)} Let  $p$ be  odd. Let $\S_N=\SS^N$.
   For any $c \ne 0$,
  the point process
$$
\PP_N\equiv\sum_{\s \in \Sigma_N} \delta \{ 2^{N}
       K_p (2/\sqrt{2\pi})
    |Y_N(\s) - c | \}
\Eq(pp2)
  $$
converges weakly to the standard Poisson point process  on $\R_+$.
\item{(ii)}   Let $p$ be odd  and $c=0$, or let $p$ be even
  and $c \neq 0$.    Denote by    $ \Sigma_N$   the  space
  of the $2^{N-1}$ equivalence classes in $\SS^N$ where $\s$ and $-\s$
  are identified.
   Then  the point process
 $$
\PP_N\equiv \sum_{\s \in  \Sigma_N } \delta \{ 2^{N-1}
    K_p (2/\sqrt{2\pi})
    |Y_N(\s) - c | \},
\Eq(pp3)
  $$
converges weakly to the standard Poisson point process  on $\R_+$. }

\noindent{\it Proof of Theorem~\thv(uthyyy).}
       We should check the assumptions of Theorem \thv(ABS.3).
  The Assumptions~A (i), for any $\eta \in ]0, p/2[$,
      and (ii) have been already
   verified in the proof of Theorem \thv(unp-not0).
  We must check (iii) and also
        the assertion  \eqv(ab.abs.5)
   on an appropriate subset $\QQ_N$.

      We will use the construction of the matrix
  $C_p(\s^1,\dots, \s^{\ell})$ explained in the proof
  of Theorem~\thv(unp-not0), see \eqv(ccb).
     Let us introduce
      the Fourier transform
$$
 f^{\s^1,\dots, \s^{\ell}}(t_1, \dots, t_\ell )=
      \E \exp \Big(i   [t_1(Y_N(\s^1)-\E Y_N(\s^1))+
    \cdots + t_k (Y_N(\s^k)-\E Y_N(\s^\ell ))] \Big).
\Eq(uu)
$$
  A simple computation shows that
 $$
f^{\s^1,\dots, \s^\ell }(t_1, \dots, t_\ell )=\prod_{m=1}^{
   {N \choose p} } \phi\Big( {N \choose p}^{-1/2}
    \{ C_p(\s^1,\dots, \s^\ell ) \vec t\}_m \Big),
  \Eq(rtt)
$$
  where   $\{ C_p(\s^1,\dots, \s^\ell ) \vec t\}_m$
  is the $m$th  coordinate of the product of the matrix
    $C_p(\s^1,\dots, \s^\ell)$ with the vector
  $\vec t =(t_1,\dots, t_\ell )$.

      Assumption A (iii) is valid due
 to the following proposition.

\proposition{\TH(pr1)} {\it There exists a constant, $Q=Q(r, \ell, b_1,\dots,
      b_\ell) >0$,
      such that, for any  $(\s^1,\dots, \s^\ell) \in \S_N^{\otimes
      \ell}$,
   any   $r\leq \ell$,  if ${ \rank }\,B_N(\s^1,\dots
\s^\ell)=r$,
$$
  \P\left( \forall_{i=1}^{\ell}:
      | Y_N(\s^i) - c  |
       \leq  \widetilde \delta_ N b_i  \right)
   \leq   [\widetilde{\delta}_N]^r Q N^{pr/2+1}.
\Eq(tra)
$$
}

\noindent{\it  Proof.}
   Recall that it follows from the hypothesis that the rank of
the matrix $C_p(\s^1,\dots, \s^\ell)$ equals $r$.
   Let us remove from this matrix $\ell-r$
     columns such that the remaining  $r$
  columns are linearly independent.  They correspond to
  a certain subset of $r$ configurations. Without loss of generality,
  we may assume that they are
    $\s^1,\dots, \s^r$, i.e., we obtain the matrix $ C_p(\s^1,\dots, \s^r)$.
     Obviously,
$$
  \P\left( \forall_{i=1}^{\ell}:
      | Y_N(\s^i) - c  |
       \leq  \widetilde \delta_ N b_i  \right)
\leq  \P\left( \forall_{j=1}^{r}:
      | Y_N(\s^j) - c  |
       \leq  \widetilde \delta_ N b_i  \right)
\Eq(ab.tra.1)
$$
   Then
$$
\eqalign{
&  \P\left( \forall_{i=1}^{\ell}:
      |  Y_N(\s^i) - c  |
       \leq  \widetilde \delta_ N b_i  \right) \cr
& \leq  \frac{1}{(2\pi)^{r}}
   \lim_{D\to \infty} \int \limits_{[-D,D]^{r}}
   \big| f_N^{\s^1,\dots, \s^r } (t_1,\dots, t_r)\big|
\prod\limits_{j=1}^r
  \frac{e^{i t_{j} b_j \widetilde \delta_N }-
      e^{-i t_{j} b_j \widetilde \delta_N } }{
     2 i t_{j }  }
  dt_{j}.
}
\Eq(tt1.1)
$$
  As $\widetilde \delta_N =O(2^{-N})$,
 the integrand in \eqv(tt1.1) is bounded by
$$
  \left|\frac{e^{i t_{j} b_j \widetilde \delta_N }-
      e^{-i t_{j} b_j \widetilde \delta_N } }{
     2 i t_{j }   }\right|
\leq  \min \left( Q_0 2^{-N},
    2|t_j|^{-1} \right),
\Eq(ab.tt1.1)
$$
  with a constant, $Q_0=Q_0(b_j)$.
 Next, let us choose in the matrix $ C_p(\s^1,\dots,\s^r)$
       any $r$ linearly independent rows and
    construct from  them an  $r\times r$
   matrix, $\bar C^{r\times r}$. Then, by \eqv(rtt) and by
   Assumption $B$   on $\phi(s)$
$$
 | f^{\s^1,\dots, \s^r}_N(\vec t)|
   \leq \prod_{j=1}^r
    \left|\phi\Big( {N \choose p}^{-1/2}
    \{  \bar C^{r\times r} \vec t \}_j\Big)\right|\leq
    \prod_{j=1}^r \min \left(1,
    \tilde Q_0 N^{p/2}
     \left|\{ \vec t\bar C^{r\times r}\}_j\right|^{-1}\right),
\Eq(ab.tt1.2)
$$
  with $\tilde Q_0>0$.
    Hence, the absolute value of the
         integral \eqv(tt1.1) is bounded by
 the sum of two terms,
$$
\eqalign{
& Q_0(b_1)\cdots Q_0(b_r) 2^{-N r}
  \int\limits_{\|\vec t\|<2^{N r } }
   \prod\limits_{j=1}^r \min
    \Big(1,  \tilde Q_0 N^{p/2}
         \big|\{ \bar C^{r\times r} \vec t \}_j
 \big|^{-1}  \Big)   dt_{j}\cr
& +
    \int\limits_{\|\vec t\|> 2^{ N r} }
         \prod_{j=1}^r ( 2 t_m^{-1})
       \prod\limits_{j=1}^r \min
    \Big(1,  \tilde Q_0 N^{p/2}
         \big|\{ \bar C^{r\times r}\vec t \}_j\big|^{-1}
\Big) dt_{j}.
}
\Eq(yyl)
$$
 Recall  that the matrix $\bar C^{r\times r}$ has matrix elements
   $\pm1$ and rank $r$. Since the total number of such matrices is at
   most
$2^{r^2}$, the smallest absolute value of the determinant of all such
   matrices is some positive number that does not depend on $N$, but
   only on $r$.
 Therefore, the change of variables, $\vec \eta=
   \bar C^{r\times r}\vec t$, in the first term
   shows that the integral over $\|\vec t\|< 2^{r N}$
  is  of  order at most
$N^{pr/2}\ln 2^{r N}\sim N^{pr/2+1}$.
        Thus the first term of \eqv(yyl) is bounded by
  $Q_1 2^{-N r} N^{pr/2+1}$, with some constant $Q_1<\infty$.
      Using the change of variables
  $\vec \eta = 2^{-r N } \vec t$ in the second term
  of \eqv(yyl), one can see  that
    the integral over $\|\vec t\|>2^{ N r}$
  is bounded by  $Q_2 2^{-N r} N^{pr/2}$,
  with some constant $Q_2<\infty$.
  This concludes the proof.
\endproof

      Finally, let us fix any
     $\eta\in ]0, 1/2[$ and
 introduce $\QQ_N=\QQ_{N, \ell, q}^{\eta/p}$
 ( defined in \eqv(qqqqq)) with
$q=(p \max_{k=0, \dots, [p/2]}{2k \choose p}(k-1)!!)^{-1}$.
     By \eqv(qqqq) and \eqv(aaaa), it is a subset of
  $\RR_{N, \ell}^{\eta}$, and
  $|\Sigma_N^{\otimes \ell} \setminus \QQ_N|$
   is smaller than $2^{N \ell} e^{- h N^{1-2\eta}}$,
  with some $h>0$.  We need to verify   \eqv(ab.abs.5) for  $\QQ_N$.
 We abbreviate
 $$
\vec{W}_N\equiv v^{-1}\big((c- \E Y_N(\s^1)),
     \dots,(c-  \E Y_N(\s^{\ell})) \big).
\Eq(w)
$$
 For any $\s^1,\dots, \s^\ell \in \QQ_{N}$,
 we  split
$$
 \P(\forall_{i=1}^{\ell} |Y_N(\s^i) -c|<b_i \widetilde \d_N )
    =\sum_{m=1}^4 I_N^m(\s^1,\dots, \s^\ell),
\Eq(op)
$$
   where
$$
\eqalign{
 I_N^1(\s^1, \dots, \s^\ell)
  &=  \int\limits_{\R^{\ell}}
   \prod_{j=1}^{\ell} \frac{ e^{i t_j b_j \widetilde \delta_N}-
    e^{i t_j b_j \widetilde \delta_N}}{2 i t_j}
   e^{i \vec t \cdot \vec{W}_N }
   e^{-\vec t B_N(\s^1, \dots, \s^\ell) \vec t /2}
   d \vec t\cr
&-
\int\limits_{\|\vec t\| \geq \epsilon N^{p/6}}
   \prod_{j=1}^{\ell} \frac{ e^{i t_j b_j \widetilde \delta_N}-
    e^{-i t_j b_j \widetilde \delta_N}}{2 i t_j}
   e^{i \vec t \cdot \vec{W}_N }
   e^{-\vec t B_N(\s^1, \dots, \s^\ell) \vec t /2}
 d \vec t,
}\Eq(11)
$$
$$
 I_N^2(\s^1, \dots, \s^\ell)
   =  \int\limits_{\|\vec t\|<\epsilon N^{p/6}}
     \prod_{j=1}^{\ell} \frac{ e^{i t_j b_j \widetilde \delta_N}-
    e^{-i t_j b_j \widetilde \delta_N}}{2 i t_j}
   e^{i \vec t \cdot \vec{W}_N }
  \big(f_N^{\s^1,\dots, \s^\ell}(\vec t)
  -  e^{-\vec t B_N(\s^1, \dots, \s^\ell) \vec t /2}\big)
    d \vec t,
\Eq(12)
$$
$$
 I_N^3(\s^1, \dots, \s^\ell )
   =
  \int\limits_{\epsilon N^{p/6}<\|\vec t\|<\delta \sqrt{N^p}}
   \prod_{j=1}^{\ell} \frac{ e^{i t_j b_j \widetilde \delta_N}-
    e^{-i t_j b_j \widetilde \delta_N}}{2 i t_j}
    e^{i \vec t \cdot \vec{W}_N }
   f_N^{\s^1,\dots, \s^\ell}(\vec t)
   d \vec t,
\Eq(13)
$$
and
 $$
 I_N^4(\s^1, \dots, \s^\ell)=
     (2\pi)^{-\ell}
    \lim_{D\to \infty}
   \int\limits_{[-D, D]^{\ell}\cap \|\vec t\|>\delta \sqrt{N^p}}
      \prod_{j=1}^{\ell} \frac{ e^{i t_j b_j \widetilde \delta_N}-
    e^{-i t_j b_j \widetilde \delta_N}}{2 i t_j}
    e^{i \vec t \cdot \vec{W}_N }
   f_N^{\s^1,\dots,\s^\ell} (\vec t)  d\vec t,
 \Eq(15)
 $$
     with some $\epsilon, \delta>0$ to be chosen later.

       The first part of  $I_N^1(\s^1,\dots, \s^\ell )$
   is exactly the quantity
  $\P(\forall_{i=1}^{\ell}: |Z_N(\s^i)- c|< b_i \widetilde
      \delta_N)$.
  Note that
   $$
\Big| \prod_{j=1}^{\ell} \frac{ e^{i t_j b_j \widetilde \delta_N}-
    e^{- t_j b_j \widetilde \delta_N}}{2 i t_j}  \Big| \leq
      Q 2^{-N \ell} ,
\Eq(stb)
$$
  with some $Q<\infty$. Then the  second part of $I_N^1$
   is  exponentially smaller than
 $2^{-\ell N}$,  for all
 $(\s^1,\dots, \s^{\ell}) \in \QQ_{N}$.
  We must show that  $I_N^2, I_N^3, I_N^4$ are
     $o(2^{-N \ell})$,  for all
 $(\s^1,\dots, \s^{\ell}) \in \QQ_{N}$.
  This  is easy due to the following  proposition.

\proposition{\TH(pr4)} {\it
  There exist  constants, $C<\infty$, $\epsilon,\theta,\delta>0$,
such that, for all
     $(\s^1, \dots, \s^\ell) \in \QQ_{N}$, the following estimates
  hold:
\item{(i)} For all $\| \vec t \|<\epsilon N^{p/6}$,
$$
\big |
 f_N^{\s^1,\dots, \s^\ell}
(\vec t )- e^{-\vec t B_N(\s^1, \dots, \s^\ell) \vec t/2}
\big|\leq \frac{ C \| \vec t\|^3}{\sqrt{N^p}}e^{-\vec t B_N(\s^1, \dots, \s^\ell)
   \vec t/2}.
\Eq(p1)
$$
\item{(ii)}      For all
$ \|\vec t\|<\delta\sqrt{N^p}$,
 $$
\big |
  f_N^{\s^1,\dots, \s^\ell}(\vec t )\big| \leq e^{-\theta
   \|\vec t\|^2}.
\Eq(p3)
$$
 }

\noindent{\it Proof.}  The proof is elementary and completely
analogous to the corresponding estimate in the
proof of the Berry-Essen inequality.
     All  details are completely analogous to those
  in the proof of Lemma~3.5 in \cite{BK2} and
  therefore are omitted.  \endproof

     Using \eqv(p1) and    \eqv(stb), we see that
   $I^2_N(\s^1,\dots,\s^l)=
    O(N^{-p/2}) 2^{-N \ell }$.
    The third term,
   $I^3_N(\s^1,\dots,\s^l)$,  is exponentially smaller
   than $2^{-N\ell}$   by \eqv(p3).

  Finally, by \eqv(stb)
    we  may estimate $  I_N^4(\s^1, \dots \s^{\ell})$ roughly as
$$
|  I_N^4(\s^1, \dots \s^{\ell})|\leq Q 2^{-\ell N}
  \int\limits_{\|\vec t\|>\delta \sqrt{N^p}}|  f_N^{\s^1,\dots,\s^k}
(\vec t)| d \vec t,
\Eq(16)
$$
  with some constant $Q<\infty$.
        By the construction of the set
  $\QQ_{N}$ \eqv(qqqqq),  for any $(\s^1,\dots,\s^\ell) \in
   \QQ_{N}$, the matrix
   $C_1(\s^1,\dots, \s^\ell)$,
   (i.e., the matrix with $N$ rows, the $k$th  row
    being $\s^1_k, \s^2_k, \dots, \s^{\ell}_k$),
      contains at least $2^{\ell-1}$ possible
  different rows, each row being present
   at least     $2^{-\ell}N(1+o(1))$ times.
   Consequently,  each of these rows is present in
       the matrix
    $C_p(\s^1,\dots, \s^\ell)$ at least   $2^{-\ell} N^p(1+o(1))$
   times,  for any $p\geq 2$.
  Then,  by \eqv(rtt),
       $f_N^{\s^1,\dots,\s^\ell}( \vec t )$
   is the product of at least $2^{\ell-1}$
         different characteristic
   functions,  each is taken
   to  the power at least $2^{-\ell} N^p(1+o(1))$.
    Let us  fix from  a set of different
  rows of $C_p(\s^1,\dots,\s^\ell)$
         $\ell$ linearly independent ones,
   and denote by $ \bar C$ the square matrix composed of them.
   Then there  exists $\zeta(\delta)>0$, such that
   $\sqrt {\vec t \bar C^T \bar C \vec t/v^2
       }\geq \zeta $, for all
    $\vec t$,  with $\|\vec t\|>\d$.
       Changing  variables
    $\vec s= {N \choose p}^{-1/2 }\bar C \vec t  $ in \eqv(16),
 one gets the bound
$$
|I_N^4(\s^1, \dots, \s^\ell )|
   \leq  Q 2^{-\ell N} N^{p \ell /2}
        \int\limits_{\|\vec s\|>\zeta }
     \prod\limits_{m=1}^\ell
       \big| \phi(s_m)
  \big|^{2^{-(\ell-1)} N^p(1+o(1))}
    ds_{m}.
\Eq(hj)
$$
  Assumption B
  made on $\phi(s)$  implies that
$\phi(s)$ is  aperiodic,  and thus
     $|\phi(s)|<1$, for any $s \ne 0$.
 Moreover, for any $\zeta>0$, there exists
  $h(\zeta)>0 $, such that
     $|\phi(s)|<1-h(\zeta)$,
    for all
  $s$ with $|s|>\zeta/\ell$. Therefore, the right-hand side of  \eqv(hj)
does not exceed
$$
Q 2^{-N \ell } N^{p \ell /2}
    (1-h(\zeta))^{2^{-(\ell-1) } N^p(1+o(1))-2}
    \int\limits_{\|\vec s\|>\eta }
     \prod\limits_{m=1}^\ell
   \big| \phi(s_m)
  \big|^{2} ds_m,
\Eq(18)
$$
 where  the integral is finite
    again due to  Assumption B.
  Therefore,
    $I_N^4(\s^1,\dots,\s^\ell)$
  is  exponentially smaller than  $ 2^{-N \ell}$.
       This concludes the proof of \eqv(ab.abs.5)
   on $\QQ_N$
  and of the theorem. \endproof

\medskip
\noindent {\bf 3.3. Short range spin glasses.}

As a final example, we consider  short-range spin glass  models.
To avoid unnecessary
complications, we will look at models on the $d$-dimensional  torus, $\L_N$,
of length $N$. We consider Hamiltonians of the form
$$
H_N(\s)\equiv - N^{-d/2}\sum_{A\subset\L_N} r_A J_A \s_A
\Eq(srsg.1)
$$
where e $\s_A\equiv \prod_{x\in A}\s_x$, $r_A$ are given constants,
and $J_A$ are random variables.
We will introduce some notation:

\item{(a)} Let $\AA_N$ denote the set of all $A\subset \L_N$, such
that $r_A\neq 0$.
\item{(b)} For any two subsets, $A,B\subset\L_N$, we say that
$A\sim B$, iff there exists $x\in\L_N$ such that $B=A+x$. We denote by
$\AA$ the set of equivalence classes of $\AA_N$ under this relation.

We will assume that the constants, $r_A$, and the random variables, $J_A$,
satisfy the following conditions:
\item{(i)} $r_A=r_{A+x}$, for any $x\in \L_N$;
\item{(ii)} there exists $k\in \N$,
such that any equivalence class in $\AA$ has a representative $A\subset\L_k$;
we will identify the set $\AA$ with a uniquely chosen set of
representatives contained in $\L_k$.
\item{(iii)} $\sum_{A\subset \L_N :} r_A^2 =N^d$.
\item{(iv)}
$J_A$, $A\in \Z^d$, are a family of independent
 random variables, such that
\item{(v)} $J_A$ and $J_{A+x}$
are identically distributed for any $x\in \Z^d$;
\item{(vi)} $\E J_A=0$ and $\E J_A^2=1$, and $\E J_A^3<\infty$;
\item{(vii)} For any $A\in\AA$, the Fourier transform
$\phi_A(s)\equiv\E \exp \left(i s J_A \right)$, of $J_A$ satisfies
 $|\phi_A(s)|=O(|s|^{-1})$ as $|s|\to  \infty$.

   Observe that $\E H_N(\s)=0$,
$$
b(\s,\s')\equiv N^{-d}\E H_N(\s)H_N(\s') =N^{-d}\sum_{A\subset
\L_N} r_A^2 \s_A\s_A'\leq 1 \Eq(srsg.2)
$$
where equality holds, if $\s=\s'$.

     Note that $Y_N(\s)=Y_N(\s')$
  (resp.  $Y_N(\s)=-Y_N(\s')$ ),
 if and only if, for all $A\in \AA_N$,
 $\s_A=\s_A'$ (resp. $\s_A=-\s_A'$).
E.g., in the standard Edwards-Anderson model,
with nearest neighbor pair interaction, if $\s_x$ differs from $\s_x'$
 on every second site,
$x$, then $Y_N(\s)=-Y_N(\s')$, and if $\s'=-\s$,  $Y_N(\s)=Y_N(\s')$.
In general, we will consider two configurations, $\s,\s'\in S^{\L_N}$,
 as equivalent, iff for all $A\in \AA_N$, $\s_A=\s_{A}'$. We denote the set
of these equivalence classes by $\S_N$. We will assume in the sequel
that there is a finite constant, $\Gamma\geq 1$, such that
$|\S_N|\geq 2^{N^d} \Gamma^{-1}$. In the special case of $c=0$, the
equivalence relation
will be extended to include the case $\s_A=-\s_{A}'$, for all $A\in
\AA_N$. In most reasonable examples (e.g. whenever nearest neighbor pair
interactions are included in the set $\AA$),
the constant $\G\leq 2$ (resp.~$\G\leq
4$, if $c=0$)).

\theo{\TH(unppp)}{\it Let $c \in \R$, and  $\Sigma_N$  be
        the space of equivalence classes defined before.
Let $
\d_N\equiv  |\Sigma_N|^{-1}e^{c^2/2}\sqrt{\frac {\pi}2}$.
  Then the point process
$$
\PP_N\equiv \sum_{\s \in \Sigma_N} \delta_{\{\d_N^{-1}
              |H_N(\s)-c|\}},
 \Eq(pll)
$$
  converges weakly to the standard Poisson point process on $\R_+$.

\noindent If, moreover, the random variables $J_A$ are Gaussian,
then, for any $c\in \R$, and $0\leq \a<1/4$, with  $ \d_N\equiv
|\Sigma_N|^{-1}e^{N^{2\a}c^2/2}\sqrt{\frac {\pi}2}$,
 the point process
$$
\PP_N\equiv \sum_{\s \in \Sigma_N} \delta_{\{\d_N^{-1}
              |H_N(\s)-cN^\a|\}},
 \Eq(pll.1)
$$
  converges weakly to the standard Poisson point process on $\R_+$.}

\noindent{\it Proof.}
We will now show that the assumptions A of Theorem \thv(ABS.3) hold.
  First, the point (i) of Assumption A is verified due to
     the following proposition.

\proposition{\TH(SRSG.0)}{\it Let $\RR_{N,\ell}^\eta$ be defined as in
\eqv(ab.abs.6). Then, in the setting above, for all $0\leq \eta<\frac 12$,
$$
|\RR_{N,\ell}^\eta|\geq |\S_N|^{\ell} \left(1- e^{- h N^{d(1-2\eta)}} \right),
\Eq(srsg.0)
$$
  with some constant $h>0$.
}

\noindent {\it Proof.}      Let  $\E_\s$ denote the expectation under the
uniform probability  measure on  $\{-1,1\}^{\L_N}$. We will show that
there exists a constant, $K>0$, such that, for any $\s'$,
and any $0\leq \d_N\leq 1$
$$
\P_{\s}(\s: b(\s,\s')>\d_N)\leq \exp (-K\d_N^2 N^d ).
\Eq(srsg.3)
$$
  Note that without loss, we can take
 $\s'\equiv 1$. We want to use the exponential Chebyshev inequality
 and thus  need to estimate the
Laplace transform
$$
\E_\s \exp\left(tN^{-d}\sum_{A\in \L_N} r_A^2 \s_A\right).
\Eq(srsg.4)
$$
Let us assume for simplicity that $N=nk$ is a multiple of $k$,
and  introduce the sub-lattice, $\L_{N,k}\equiv \{0,k,\dots,(n-1)k,nk\}^d$. Write
$$
\sum_{A\in \L_N} r_A^2 \s_A =\sum_{A\in \AA} \sum_{y\in \L_{N,k}}
\sum_{x\in \L_k} r_{A+y+x}^2\s_{A+y+x}
\equiv\sum_{x\in \L_k} Z_x(\s)
\Eq(srsg.5)
 $$
where
$$
 Z_x(\s)=\sum_{y\in \L_{N,k}} Y_{y,x}(\s)
\Eq(srsg.6)
$$
has the nice feature that, for fixed $x$, the summands
$$
Y_{x,y}(\s)\equiv \sum_{A\in \AA} r_{A+y+x}^2\s_{A+y+x}
$$
are independent for different $y, y'\in \L_{n,k}$ (since the sets  $A+y+x$ and
$A'+y'+x$ are disjoint for any $A,A'\in \L_k$).
Using the H\"older inequality repeatedly,
$$
\eqalign{ \E_\s\exp\left(t\sum_{x\in \L_k} Z_x(\s)\right) &\leq
\prod_{x\in \L_k} \left[\E_\s e^{t k^d Z_x(\s)}\right]^{k^{-d}}
\cr &=
 \prod_{x\in \L_k}\prod_{y\in \L_{N,k}} \left[\E_\s e^{tk^d Y_{x,y}(\s)}
\right]^{k^{-d}}
\cr
&=\left[\E_\s e^{tk^d Y_{0,0}(\s)}\right]^{N^dk^{-d}}
}
\Eq(srsg.7)
$$
It remains to estimate the Laplace transform of
$Y_{0,0}(\s)$,
$$
\E_\s \exp\left({tk^d Y_{0,0}(\s)}\right)
=\E_\s \left({tk^d\sum_{A\in \L_k} r_{A}^2\s_{A}}\right),
\Eq(srsg.8)
$$
 and, since $\E_\s\s_A=0$, using that $e^x\leq 1+x+\frac {x^2}2 e^{|x|}$,
$$
\E_\s\exp\left({tk^d\sum_{A\in \L_k} r_{A}^2\s_{A}}\right) \leq
\E_\s \exp\left({\frac {t^2}2 k^{2d} \left(\sum_{A\in \L_k}
r^2_A\right)e^{tk^d
 \sum_{A\in \L_k} r^2_A}}\right)
\equiv \E_\s \exp\left({\frac {t^2}2 C e^{t D}}\right),
\Eq(srsg.9)
$$
so that
$$
\E_\s\exp\left(tN^{-d}\sum_{x\in \L_k} Z_x(\s)\right) \leq
\exp\left({ {N^{-d}} \frac {t^2}2 C' e^{N^{-d} t D}}\right),
\Eq(srsg.10)
$$
with constants, $C,C',D$, that do not depend on $N$.
To conclude the proof of the lemma, the exponential Chebyshev
inequality gives,
$$
\P_\s\left[b(\s,\s')>\d_N\right]\leq \exp\left({-\d_N t + {N^{-d}}
\frac {t^2}2 C' e^{tN^{-d} D}}\right). \Eq(srsg.11)
$$
Choosing $t=\e N^d\d_N $, this gives
$$
P_\s\left[b(\s,\s')>\d_N\right]\leq \exp\left({-\e\d^2_N  {N^{d}}\left(
1-\e C' e^{\e\d_N D}/2\right)}\right)
\Eq(srgs.12)
$$
 Choosing $\e$ small enough, but independent of $N$, we
obtain the assertion of the lemma. \endproof

     To verify Assumptions A (ii) and (iii), we need to introduce
 the  matrix $C=C(\s^1,\dots, \s^{\ell})$
 with $\ell$ columns and $ |\AA_N| $  rows,
        indexed
  by the subsets $A \in \AA_N$: the elements
  of each of its column are  $r_A  \s_A^1,
         r_A  \s_A^2,\dots, r_A  \s_A^\ell$,
   so that $C^T C$ is the covariance matrix, $B_N(\s^1,\dots, \s^\ell)$,
   up to a multiplicative factor $N^{d}$.

     The assumption (ii) is verified due to
   Proposition~\thv(prs). In fact,
       let us reduce $C$ to the matrix
    $\tilde C=\tilde C(\s^1,\dots, \s^{\ell})$
   with columns $\s_A^1,
      \s_A^2,\dots, \s_A^\ell$, without the constants $r_A$.
      Then, exactly as in the
  case of $p$-spin SK models, by  Proposition \eqv(prs), for any
   $(\s^1,\dots, \s^{\ell}) \in \LL_{N^d, r}^{\ell}$
    the matrix $\tilde C(\s^1,\dots,
   \s^{\ell})$ can  contain at most $2^r-1$ different columns.
   Hence, $|\LL_{N^d, r}^{\ell}|=O((2^r-1)^{N^d})$
   while $|\Sigma_N|^r \geq (2^{N^d}/\G)^r$.

       The assumption (iii) is verified as well, and
  its proof is completely analogous to that of Proposition~\thv(pr1).
  The key observation is that, again,   the number of possible non-degenerate
  matrices  $\bar C^{r\times r}$ that can be obtained from
  $C_p(\s^1,\dots,\s^\ell)$ is independent
  of $N$. But this is true since, by assumption,   the number different constants
  $r_A$ is $N$-independent.

     Finally, we define $\QQ_N$ as follows.
 For any  $A \in \A$, let
$$
\QQ_{N, \ell}^{\eta, A} =
 \Big\{(\s^1,\dots, \s^{\ell}): \forall_{1\leq i<j\leq \ell}\,\,
   r_A^2 \sum_{x \in \Z^d: x+A \subset \Lambda_N }
      \s_A^i \s_A^j < |\AA|^{-1}N^{-\eta}\Big\}.
 \Eq(aba)
$$
Let us define $\QQ_N= \bigcap_{A \in \A}
   \QQ_{N, \ell}^{\eta, A} \subset  \RR_{N, \ell}^{\eta}$.
        By Proposition \eqv(SRSG.0), applied
  to a model where $|\A|=1$, for
  any $A \in \A$, we have
 $|\SS_N^{\otimes \ell}\setminus
      \QQ_{N, \ell}^{\eta, A}| \leq 2^{N^d}\exp(-h_A N^{d(1-2\eta)})$,
  with some $h_A>0$.
    Hence, $|\RR_{N, \ell}^{\eta} \setminus \QQ_N|$
  has  cardinality smaller than $|\Sigma_N|^{\ell} \exp(-h N^{d(1-2\eta)})$,
 with some $h>0$.
     The verification of \eqv(ab.abs.5) on $\QQ_N$
 is analogous to the one in Theorem \thv(uthyyy),
  using the analogue of Proposition \eqv(pr4).
  We only note a small difference
 in the analysis of the term $I_N^4$ where
       we use the explicit construction of $\QQ_N$.
  We represent the corresponding generating function
 as the product of $|\AA |$ terms
 over different equivalence classes of $\AA$,
  with representatives $A \subset \L_k$, each term being
  $\prod_{x \in \Z^d: x+A \in \Lambda_N}\phi(N^{-d/2} r_A
    (t_1\s^1_{x+A}+\cdots +t_{\ell}\s^{\ell}_{x+A}))$.
   Next, we  use the fact that for any
  $(\s^1,\dots, \s^{\ell}) \in \QQ_N$
      each of these $|\AA|$ terms is a product of
   at  least $2^{\ell}-1$ (and of coarse at most $2^\ell $)
              different terms,  each is
   taken to the power $|\AA|^{-1}N^{d}2^{-\ell}(1+o(1))$.
This proves the first assertion of the theorem.

The proof of the second assertion, i.e., the case $\a>0$ with Gaussian
variables
$J_A$ is immediate from the estimates above and the abstract Theorem
\thv(ABS.1), in view of the fact that the condition \eqv(ab.abs.5) is
trivially verified.
\endproof

\bigskip
{\headline={\ifodd\pageno\rightheadline \else \leftheadline \fi}}
\def\rightheadline{\it  {Mertens}\hfil\tenrm\folio}
\def\leftheadline{\tenrm \folio \hfil\it  {References}}

\Refs

\item{[BFM]}
H.~Bauke, S.~Franz, and St.~Mertens.
  Number partitioning as random energy model.
  {\it Journal of Statistical Mechanics: Theory and Experiment}, page
  P04003, 2004.

\item{[BaMe]}
H.~Bauke and St. Mertens.
  Universality in the level statistics of disordered systems.
  {\it Phys. Rev. E}, 70:025102(R), 2004.

\item{[BCP]}
C.~Borgs, J.~Chayes, and B.~Pittel.
  Phase transition and finite-size scaling for the integer partitioning
  problem.
  {\it Random Structures Algorithms}, 19(3-4):247--288, 2001.

\item{[BCMP]}
C.~Borgs, J.~T. Chayes, S.~Mertens, and B.~Pittel.
 Phase diagram for the constrained integer partitioning problem.
{\it Random Structures Algorithms}, 24(3):315--380, 2004.

\item{[BCMN]}
C.~Borgs, J.~T. Chayes, S.~Mertens, and Ch.~Nair.
Proof of the local REM conjecture for number partitioning.
preprint 2005.

\item{[Bo]} A.~Bovier. Statistical mechanics of disordered
 systems. {\it
Cambridge University Press}, to appear (2005).

\item{[BK1]}
A.~Bovier and I.~Kurkova.
  Derrida's generalised random energy models. {I}. {M}odels with
  finitely many hierarchies.
  {\it Ann. Inst. H. Poincar\'e Probab. Statist.}, 40(4):439--480,
  2004.

\item{[BK2]}
A.~Bovier and I.~Kurkova.
  Poisson convergence in the restricted $k$-partioning problem.
  preprint 964, WIAS, 2004.

\item{[BKL]} A.~Bovier, I.~Kurkov, and M. L\"owe.
 Fluctuations of the free energy in the REM and the $p$-spin SK models.
{\it Ann. Probab.} 30  605--651, 2002.

%\item{[DV]} D.J.~Daley, D.~Vere-Jones, {\it An introduction
%     to the Theory of Point Processes.} Springer Series in
% Statistics, Springer-Verlag (1988).

\item{[Der1]}
B.~Derrida.
 Random-energy model: an exactly solvable model of disordered systems.
 {\it Phys. Rev. B (3)}, 24(5):2613--2626, 1981.

\item{[Der2]}
B.~Derrida.
 A generalisation of the random energy model that includes
  correlations between the energies.
 {\it J. Phys. Lett.}, 46:401--407, 1985.

\item{[Mer1]}
St. Mertens.
  Phase transition in the number partitioning problem.
  {\it Phys. Rev. Lett.}, 81(20):4281--4284, 1998.

\item{[Mer2]}
St. Mertens.
  A physicist's approach to number partitioning.
  {\it Theoret. Comput. Sci.}, 265(1-2):79--108, 2001

\endRefs

\end